\documentclass[a4paper,12pt]{article}
\usepackage{theorem}
\usepackage{latexsym,amssymb,amsfonts,amsmath}
\usepackage[dvips]{graphicx}
\newtheorem{theorem}{Theorem}
\newtheorem{lemma}{Lemma}
\newtheorem{corollary}[theorem]{Corollary}

\newtheorem{remark}[theorem]{Remark}
\newtheorem{definition}{Definition}

\newcommand{\bs}[1]{\boldsymbol{#1}}

\title{{\Large {\bf Localization of the Grover walks on spidernets and free Meixner laws  
}
}}

\author{ 
{\small 
Norio Konno,$^{1}$ 
\footnote{konno@ynu.ac.jp 
}\quad 
Nobuaki Obata,$^{2}$ 
\footnote{obata@math.is.tohoku.ac.jp 
}\quad  
Etsuo Segawa,$^{3}$ 
\footnote{To whom correspondence should be addressed. 
e-segawa@m.tohoku.ac.jp 
} 
}\\ 
{\scriptsize $^{1}$ 
Department of Applied Mathematics, Faculty of Engineering, Yokohama National University
}\\
{\scriptsize Yokohama 240-8501, Japan
} \\
{\scriptsize $^{2,3}$ 
Graduate School of Information Sciences, Tohoku University, 
}\\
{\scriptsize Sendai 980-8579, Japan. 
} 
} 

\date{\empty }
\begin{document}
\maketitle

\par\noindent
\begin{small}
\par\noindent
{\bf Abstract}. 
A spidernet is a graph obtained by adding large cycles to an almost regular tree
and considered as an example having intermediate properties of lattices and trees
in the study of discrete-time quantum walks on graphs. 
We introduce the Grover walk on a spidernet and its one-dimensional reduction.
We derive an integral representation of the $n$-step transition amplitude
in terms of the free Meixner law which appears as the spectral distribution.
As an application we determine the class of spidernets which exhibit localization.
Our method is based on quantum probabilistic spectral analysis of graphs.

\footnote[0]{
{\it Key words and phrases. Quantum walk, spidernet, and free Meixner law} 
}

\end{small}

\setcounter{equation}{0}
\section{Introduction}
\label{intro}

The study of quantum walks, tracing back to \cite{Gudder,Meyer},
has been accelerated from various aspects during the last decade,
see e.g., \cite{Ambainis,KFCSWAMW,KonnoBook,Venegas} and references cited therein.
From a mathematical viewpoint sharp contrast
between quantum walks and random walks is of particular importance.
For example, the ballistic spreading is observed in a wide class of
quantum walks \cite{ABNVW,CHKS,IKS,Konno2002,Konno2005,KLS,Katori},
i.e., the speed of a quantum walker's spreading is proportional to the time $n$
while the typical scale for a random walk is $\sqrt{n}$.
Moreover, the limit distributions of quantum walks are obtained \cite{CHKS,IKS,Konno2002,Konno2005,KLS,TS,Katori}
with a significant contrast with the normal Gaussian law in the case of random walks.
In this paper we focus on the phenomenon called \textit{localization},
which is also considered as a typical property of quantum walks, 
see \cite{CGMV2,IKS,KS} among others.
We introduce the Grover walk on a particular infinite graph called a \textit{spidernet},
consider an isotropic initial state,
and determine the class of spidernets which exhibits localization.
A spidernet is not only a new example for the localization
but also is expected to be a clue to understand localization from graph structure.
Our method is based on quantum probabilistic spectral analysis of graphs \cite{HO}.

A spidernet is obtained by adding large cycles to an almost regular tree,
see Subsection \ref{subsec:Spidernets} for definition and
see Fig.~\ref{fig:illestration of spidernets} for illustration.
It is expected to have intermediate properties between
trees and lattices,
and its spectral properties have been studied to some extent, 
see e.g., \cite{IO} for the spectral distribution of the adjacency matrix
and \cite{Urakawa2003} for estimates of the Cheeger constant and Green kernel 
in terms of spectra.
Then the standard application of the Karlin-McGregor formula
(see e.g., \cite{Obata2012}) 
yields an explicit formula for the $n$-step transition probability of the isotropic random walk on a spidernet,
where the free Meixner law appears as the spectral distribution.
This argument is along a natural extension of the result on random walks 
on a homogeneous tree due to Kesten \cite{Kesten}.
Our attempt in this paper is to establish the quantum counterpart.

In Section~\ref{sec:Grover Walks on Graphs} we introduce the Grover walk on a general graph
after the standard literatures, see e.g., \cite{Ambainis,Watrous}.
Then we formulate two concepts of localization,
that is, \textit{initial point localization} and 
\textit{exponential localization}.
Several quantum walks are known to exhibit the localization,
see e.g., \cite{CGMV2,CHKS,IKS,KLS,KS,Katori}.
For relevant discussion see also \cite{S}.

In Section~\ref{sec:Main Results} we introduce the spidernet $S(a,b,c)$ and
mention the main results.
We first obtain the integral representation of the $n$-step transition amplitude
for the Grover walk on a spidernet:
\begin{equation}\label{1eqn:main expression of probability amplitude}
\langle \bs{\psi}_0^+, U^n \bs{\psi}_0^+\rangle
=\int_{-1}^{1} \cos n\theta \,\mu(d\lambda),
\quad
n=0,\pm1,\pm2,\dots,
\end{equation}
where $\lambda=\cos\theta$ and $\mu$ is the free Meixner law of which the parameters
are determined by $a,b,c$ of the spidernet under consideration,
see Theorem \ref{mainthm:probability amplitude} for the precise statement.
The free Meixner law is a probability distribution on $[-1,1]$
which is the sum of absolutely continuous part and at most two point masses.
It is then rather easy to derive from \eqref{1eqn:main expression of probability amplitude}
the asymptotic behavior of the transition amplitude as $n\rightarrow\infty$.
In fact, only the effect of the point masses remains in the limit
and the asymptotic results follow.
In particular, we prove that the initial point localization occurs
if and only if $b>c+\sqrt{c}$, see Theorem \ref{3thm:localization for spidernet} for details.

It is instructive to consider the family of spidernets $S(\kappa,\kappa+2,\kappa-1)$,
$\kappa\ge2$.
These are obtained by suitably adding a large cycle to the homogeneous tree with degree $\kappa$.
We see from Theorem \ref{3thm:localization for spidernet} that
the initial point localization occurs for $2\leq\kappa< 10$
and no initial point localization occurs for $\kappa\ge10$
(Corollaries \ref{3cor:k<10} and \ref{3thm:estimate kappa ge10}).
While, Corollary \ref{tree} asserts that no initial point localization occurs
on a homogeneous tree either.
In the recent work \cite{Katori} we know that the Grover walk on two-dimensional lattice exhibits
the initial point localization.
These results suggest the effect of cycles for the localization of the Grover walk.

In Section \ref{sec:One-dimensional reduction} we introduce the one-dimensional
reduction of our Grover walk, called a $(p,q)$-quantum walk on $\mathbb{Z}_+$.
We determine the eigenvalues of the $(p,q)$-quantum walk with space cutoff
by extending the quantum probabilistic method together with theory of Jacobi matrices.

In Section \ref{sec:Proofs of main results} we obtain the integral expression of
the $n$-step transition amplitude of the $(p,q)$-quantum walk on $\mathbb{Z}_+$
(Theorem \ref{5thm:integral representation of four amplitudes})
and the asymptotic behavior of the transition amplitude
(Theorem \ref{5thm:localization criteria}).
With these preparations we prove the main results.

In Appendix we recall the definition of the free Meixner law
and derive the associated orthogonal polynomials.
The explicit form of the orthogonal polynomials is used to derive 
the somehow amazing result (Lemma \ref{5lem:special value}) 
which plays a key role in deriving the exponential localization.

Finally, we mention some relevant works.
The so-called CGMV method \cite{CGMV,CGMV2,GVWW,KS} is also based on the spectral analysis on the unit circle
and seems to have close connection with our approach. 
The technique to get the eigensystem of some class of quantum walks on a finite system including the Grover walk 
is established in \cite{Sze}. 
Our result is an extension to an infinite system, where the orthogonal polynomials 
with respect to the free Meixner play a key role.
Conservation of probability is an interesting question for a quantum walk,
see e.g., \cite{CHKS,IKS,KLS,SK,Katori}. 
The quantum walks studied in \cite{CHKS,IKS,KLS} are non-conservative
and the ``missing probability" is found through the weak convergence theorem
in such a way that the limit distribution is a convex combination of 
a point mass at the origin corresponding to localization and the Konno density function 
\cite{Konno2002,Konno2005} coming from ballistic spreading, see \cite{KLS} for details. 
It is not yet checked whether our Grover walks are conservative or not.
There is a large number of literatures under the name of quantum graphs, 
see e.g., \cite{GS2006} and references cited therein,
which are expected to have a profound relation to quantum walks but not yet very clear.

\section{Grover Walks on Graphs}
\label{sec:Grover Walks on Graphs}

Let $G$ be a graph with vertex set $V=V(G)$ and edge set $E=E(G)$,
i.e., $V$ is a non-empty (finite or infinite) set 
and $E$ is a subset of $\{\{u,v\}\,;\, u,v\in V, u\neq v\}$.
We often write $u\sim v$ for $\{u,v\}\in E$. 
Throughout the paper a graph is always assumed to be \textit{locally finite}, 
i.e., $\deg(u)=|\{v\in V\,;\, v\sim u\}| <\infty$ for all $u\in V$,
and \textit{connected}, i.e., every pair of vertices are connected by a walk.
An ordered pair $(u,v)\in V\times V$ is called 
a \textit{half-edge} extending from $u$ to $v$ if $u\sim v$.
Let $A(G)$ denote the set of half edges of $G$.

The state space of our Grover walk will be given by the Hilbert space
$\mathcal{H}=\mathcal{H}(G)=\ell^2(A(G))$ of square-summable functions on $A(G)$.
The inner product is defined by
\[
\langle \bs{\phi},\bs{\psi}\rangle
=\sum_{(u,v)\in A(G)} \overline{\bs{\phi}(u,v)}\,\bs{\psi}(u,v),
\qquad
\bs{\phi},\bs{\psi}\in \mathcal{H}.
\]
In general, a unit vector in $\mathcal{H}$ is called a \textit{state}.
The canonical orthonormal basis is denoted by 
$\{\bs{\delta}_{(u,v)}\,;\, (u,v)\in A(G)\}$.
For $u\in V$ let $\mathcal{H}_u$ be the closed subspace spanned by
$\{\bs{\delta}_{(u,v)}\,;\, v\sim u\}$.
Obviously, we have $\dim \mathcal{H}_u=\deg(u)$ and
the orthogonal decomposition:
\[
\mathcal{H}=\sum_{u\in V} \oplus \mathcal{H}_u\,.
\]

We next introduce unitary operators on $\mathcal{H}$.
With each $u\in V$ we associate a \textit{Grover operator} $H^{(u)}$ on $\mathcal{H}_u$
defined by means of the actions on the orthonormal basis
$\{\bs{\delta}_{(u,v)}\,;\, v\sim u\}$:
\begin{equation}
(H^{(u)})_{vw}
\equiv \langle \bs{\delta}_{(u,v)}, H^{(u)}\bs{\delta}_{(u,w)}\rangle
=\frac{2}{\deg(u)}-\delta_{vw}\,.
\end{equation} 
As is easily verified,
the Grover operator $H^{(u)}$ is a real symmetric, unitary operator on $\mathcal{H}_u$.
Then the \textit{coin flip operator} $C$ on $\mathcal{H}$ is defined by
\begin{equation}
C\bs{\delta}_{(u,v)}
=\sum_{w\sim u}(H^{(u)})_{vw}\bs{\delta}_{(u,w)}.
\end{equation}
The \textit{shift operator} $S$ is defined by
\[
S\bs{\delta}_{(u,v)}=\bs{\delta}_{(v,u)}\,.
\]
Note that $C^2=S^2=I$ (the identity operator).
Since both $C$ and $S$ are unitary operators on $\mathcal{H}$,
so is  
\[
U=SC,
\]
which is called the \textit{Grover walk} on the graph $G$.

The time evolution of the Grover walk with an initial state $\bs{\Phi}_0\in\mathcal{H}=\ell^2(A(G))$
is given by the sequence of unit vectors:
\[
\bs{\Phi}_n=U^n\bs{\Phi}_0\,,
\qquad n=0,1,2,\dots.
\]
Since $U^n$ is unitary, we have
\[
1=\|\bs{\Phi}_n\|^2=\sum_{u\in V}\sum_{v\sim u}|\bs{\Phi}_n(u,v)|^2,
\qquad n=0,1,2,\dots.
\]
Therefore, the function
\[
u\mapsto \sum_{v\sim u}|\bs{\Phi}_n(u,v)|^2,
\qquad u\in V,
\]
defines a probability distribution on $V$,
which is interpreted as the probability
of finding a Grover walker at $u\in V$ at time $n$.
Following convention we write
\begin{equation}\label{2eqn:distribution of GW}
P(X_n=u)=\sum_{v\sim u}|\bs{\Phi}_n(u,v)|^2,
\qquad u\in V.
\end{equation}
It is noted, however, that
$X_n$ is merely defined as a random variable for each $n$.
It is an interesting question to construct a discrete-time stochastic process
$\{X_n\,;\, n=0,1,2,\dots\}$ with state space $V$ reasonably reflecting 
probabilistic properties of the Grover walk.
The quantity $\bs{\Phi}_n(u,v)=\langle\bs{\delta}_{(u,v)},U^n\bs{\Phi}_0\rangle$
appearing in \eqref{2eqn:distribution of GW},
or more generally $\langle\bs{\Phi},U^n\bs{\Phi}_0\rangle$ for 
two states $\bs{\Phi},\bs{\Phi}_0$ is
called a \textit{transition amplitude}.
This is a quantum counterpart of transition probability of a Markov chain.

Since the sequence $\{P(X_n=u)\,;\, n=0,1,2,\dots\}$ defined in 
\eqref{2eqn:distribution of GW} is oscillating in general,
it is essential to study the time average:
\[ 
\overline{q}^{(\infty)}(u)
=\lim_{N\to\infty}\frac{1}{N}\sum_{n=0}^{N-1}P(X_n=u),
\qquad u\in V,
\]
when the limit exists.
For a state $\bs{\Phi}\in\mathcal{H}=\ell^2(A(G))$ we denote by
$\mathrm{supp\,}\bs{\Phi}$ the set of vertices $u\in V$
such that $\bs{\Phi}(u,v)=\langle \bs{\delta}_{(u,v)}, \bs{\Phi}\rangle\neq0$
for some $v\sim u$.

\begin{definition}[Initial point localization]
\label{def:initial point localization}
\normalfont\rm
Let $o\in V$ be a distinguished vertex
and $\bs{\Phi}_0\in\mathcal{H}=\ell^2(A(G))$ a state with $\mathrm{supp\,}\bs{\Phi}=\{o\}$.
We say that the Grover walk on $G$ with an initial state $\bs{\Phi}_0$ exhibits
\textit{initial point localization} if $\overline{q}^{(\infty)}(o)>0$.
\end{definition}

\begin{definition}[Exponential localization]
\label{def:exponential localization}
\normalfont\rm
Let $o\in V$ and $\bs{\Phi}_0$ be the same as in Definition \ref{def:initial point localization}.
We say that the Grover walk with an initial state $\bs{\Phi}_0$ exhibits
\textit{exponential localization} if there exist constant numbers $C>0$ and $0<r<1$ such that
\begin{equation}\label{2eqn:exponential localization}
\overline{q}^{(\infty)}(u)\ge Cr^{\partial(o,u)},
\qquad u\in V,
\end{equation}
where $\partial(o,u)$ stands for the graph distance between $o$ and $u$,
i.e., the length of the shortest path connecting them.
\end{definition}

\begin{remark}
In some literatures, e.g., \cite{Ahlbrecht,Joye}, 
``exponential localization" is defined when the opposite inequality
$\overline{q}^{(\infty)}(u)\le Cr^{\partial(o,u)}$ is satisfied
instead of \eqref{2eqn:exponential localization}.
This concept is more likely referred to as ``exponentially bounded" or ``exponential decay"
and is used for a different purpose.
\end{remark}

Note that the concepts of localization in Definitions 
\ref{def:initial point localization} and 
\ref{def:exponential localization} depend on the choice of an initial state.

\section{Main Results}
\label{sec:Main Results}

\subsection{Spidernets}
\label{subsec:Spidernets}

Let $G=(V,E)$ be a (locally finite and connected) graph with a distinguished vertex $o\in V$.
We introduce a stratification of $G$ by
\[
V=\bigcup_{j=0}^\infty V_j\,,
\qquad V_j=\{u\in V\,;\, \partial(u,o)=j \}.
\]
Then for $\epsilon\in\{+,-,\circ\}$ we define a function $\omega_\epsilon$ on $V$ by
\[ 
\omega_\epsilon(u)
=|\{v\in V_{j+\epsilon}\,;\, v \sim u\}|,
\qquad u\in V_j\,,
\]
where we understand $j+\epsilon=j+1,j-1,j$ for $\epsilon=+,-,\circ$, respectively.
Note that $\deg(u)=\omega_+(u)+\omega_\circ(u)+\omega_-(u)$ for $u\in V$.

A graph is called a \textit{spidernet}
if there exist a distinguished vertex $o\in V$ and integers $a,b,c$ with
\begin{equation}\label{3eqn:constraint abc}
a\ge1, \qquad b\ge2, \qquad 1\le c\le b-1
\end{equation}
such that
\begin{gather*}
\omega_+(u)=
\begin{cases}
	a, & \text{for  $u=o$}, \\ 
    c, & \text{otherwise},
\end{cases}
\quad
\omega_-(u)=
\begin{cases}
	0, & \text{for $u=o$}, \\
    1, & \text{otherwise},
\end{cases}
\\
\omega_\circ(u)=
\begin{cases}
	0, & \text{for $u=o$}, \\ 
    b-c-1, & \text{otherwise}.
\end{cases}
\end{gather*}
Such a spidernet is denoted by $S(a,b,c)$.
It is noted that $S(a,b,c)$ is not necessarily determined uniquely by the parameters $a,b,c$.
By definition we have
\begin{equation}
\deg(u)=
\begin{cases}
	a, & \text{for $u=o$}, \\ 
    b, & \text{otherwise},
\end{cases}
\end{equation}
and 
\[
|V_0|=1, 
\qquad
|V_j|=ac^{j-1}\,,\quad j=1,2,\dots.
\]
Hence a spidernet is an infinite graph.

\begin{figure}[htbp]
 \begin{center}
  \includegraphics[width=300pt]{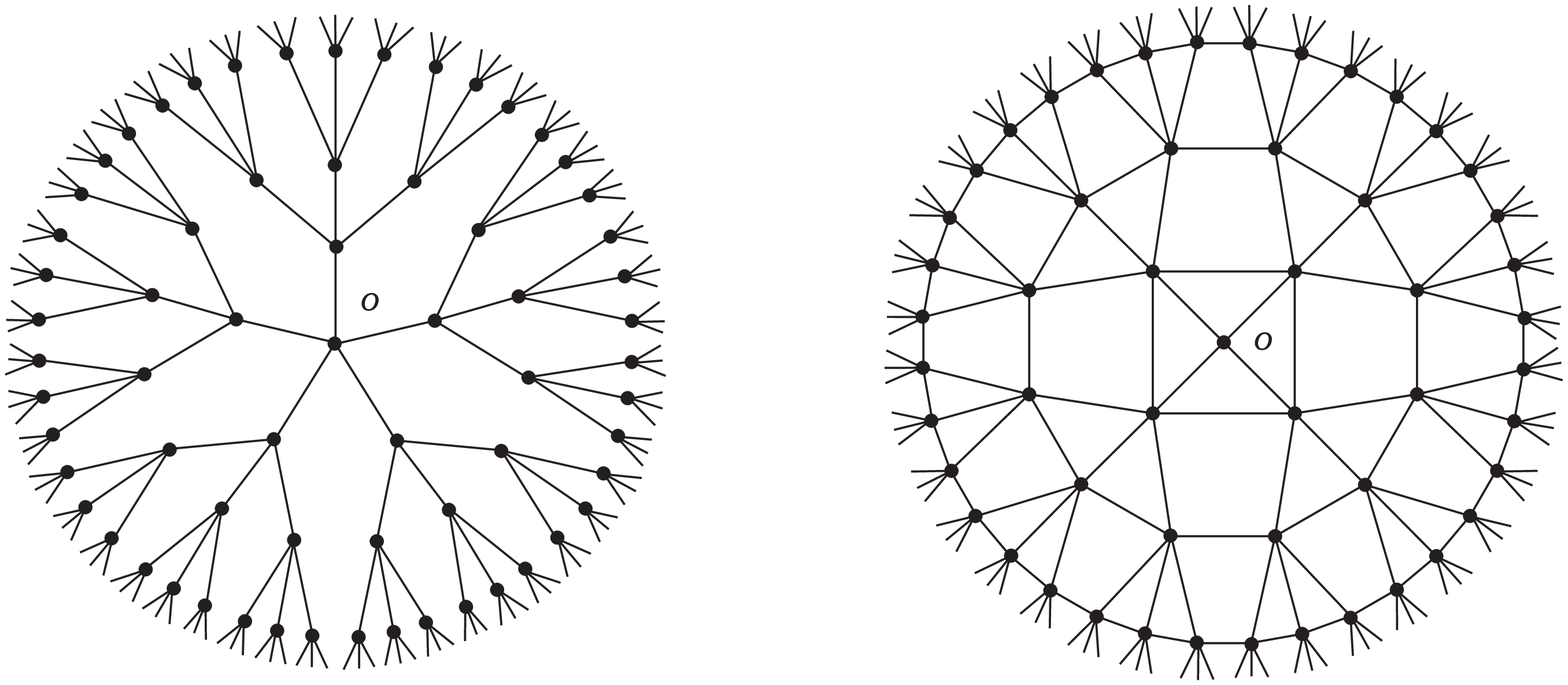}
 \end{center}
\caption{$S(5,4,3)$ and $S(4,6,3)$}
\label{fig:illestration of spidernets}
\end{figure}

A spidernet $S(a,b,b-1)$ is a tree.
In particular, $S(\kappa,\kappa,\kappa-1)$ with $\kappa\ge2$ is the homogeneous tree of degree $\kappa$.
While, a spidernet $S(\kappa,\kappa+2,\kappa-1)$ is obtained by adding a large cycle to each stratum
of the homogeneous tree of degree $\kappa$.
A typical example is shown in Fig.~\ref{fig:illestration of spidernets}; however,
note that a spidernet $S(\kappa,\kappa+2,\kappa-1)$ is not uniquely determined by
the parameter $\kappa\ge2$.
\subsection{Grover walks on spidernets}
We focus on the Grover walk $U$ on a spidernet $G=S(a,b,c)$.
Recall that the state space is given by $\mathcal{H}=\ell^2(A(G))$, 
of which the canonical orthonormal basis is denoted by $\{\bs{\delta}_{(u,v)}\,;\, (u,v)\in A(G)\}$.
Define a state $\bs{\psi}^+_0\in\mathcal{H}$ by
\begin{equation}\label{3eqn:initial state}
\bs{\psi}^+_0
= \frac{1}{\sqrt{a}} \sum_{v\sim o} \bs{\delta}_{(o,v)}\,,
\end{equation}
which is taken to be the initial state of our Grover walk.
Note that $\bs{\psi}^+_0$ is characterized by
$\mathrm{supp\,}\bs{\psi}^+_0=\{o\}$ and being isotropic.

We now list the main results of this paper.

\begin{theorem}[Integral representation of transition amplitude]
\label{mainthm:probability amplitude}
Let $U$ be the Grover walk on a spidernet $S(a,b,c)$ with an initial state $\bs{\psi}^+_0$
defined by \eqref{3eqn:initial state}.
Let $\mu$ be the free Meixner law with parameters $q,pq,r$,
where  
\begin{equation}\label{3eqn:p and q from spidernet}
p=\frac{c}{b}\,,
\qquad
q=\frac{1}{b}\,,
\qquad
r=\frac{b-c-1}{b}\,.
\end{equation}
Then for all $n=0,\pm1,\pm2,\dots$ it holds that
\begin{equation}\label{3eqn:main expression of probability amplitude}
\langle \bs{\psi}_0^+, U^n \bs{\psi}_0^+\rangle
=\int_{-1}^{1} \cos n\theta \,\mu(d\lambda),
\quad
\lambda=\cos\theta.
\end{equation}
\end{theorem}

For the definition of the free Meixner law, see Appendix.
It is also noted that \eqref{3eqn:main expression of probability amplitude} admits
an alternative expression:
\[
\langle \bs{\psi}_0^+, U^n \bs{\psi}_0^+\rangle
=\int_{-1}^{1} T_{|n|}(\lambda) \,\mu(d\lambda),
\quad
n=0,\pm1,\pm2,\dots,
\]
where $T_n$ is the \textit{Chebyshev polynomials of the first kind} defined by
\[
\cos n\theta=T_n(\cos\theta),
\qquad n=0,1,2,\dots,
\]
see e.g., \cite{Chihara,HO}.

From Theorem \ref{mainthm:probability amplitude} we will derive some results 
on initial point localization of our Grover walks.
In fact, we determine the class of spidernets $S(a,b,c)$ which exhibit
initial point localization as follows.

\begin{theorem}\label{3thm:localization for spidernet}
Let $U$ be the Grover walk on a spidernet $S(a,b,c)$ with an initial state $\bs{\psi}^+_0$
defined by \eqref{3eqn:initial state}.
It then holds that
\[
\langle \bs{\psi}_0^+, U^n \bs{\psi}_0^+\rangle
\sim
w \cos n \tilde\theta,
\qquad
\text{as $n\rightarrow\infty$},
\]
where
\[
w=\max\left\{\frac{(b-c)^2-c}{(b-c)(b-c+1)}\,,0\right\},
\quad
\cos\tilde\theta=-\frac{1}{b-c}\,,
\quad
0<\tilde\theta<\pi.
\]
In particular, if $b>c+\sqrt{c}$, then the initial point localization occurs:
\[
\overline{q}^{(\infty)}(o)
=\lim_{N\rightarrow\infty}\frac1N\sum_{n=0}^{N-1} P(X_n=o)
=\frac{w^2}{2}>0.
\]
If $b\le c+\sqrt{c}$, then no localization occurs:
\[
\lim_{n\rightarrow\infty}P(X_n=o)=0
\quad{and}\quad
\overline{q}^{(\infty)}(o)=0.
\]
\end{theorem}

It is instructive to consider the family of spidernets $S(\kappa,\kappa+2,\kappa-1)$, $\kappa\ge2$.
Note that $S(\kappa,\kappa+2,\kappa-1)$ is obtained by adding a large cycle to each stratum
of $S(\kappa,\kappa,\kappa-1)$, which is the homogeneous tree of degree $\kappa$.
Below we list some results obtained immediately from Theorem \ref{3thm:localization for spidernet}.

\begin{figure}
\begin{center}
\includegraphics[width=225pt]{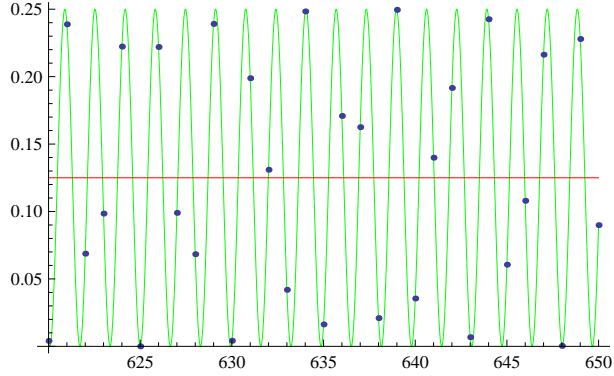}
\caption{The Grover walk on $S(4,6,3)$ from time $n=620$ to $n=650$
(see Corollary \ref{3cor:k<10}): 
The dots stand for $P(X_n=o)$ calculated by numerical simulation. 
The curve is the graph of $(1/4)\cos^2(t\tilde\theta)$ 
and the horizontal line depicts the time averaged limit probability $\overline{q}^{(\infty)}(o)$.}
\label{fig:one}      
\end{center}
\end{figure}

\begin{corollary}\label{3cor:k<10}
Let $2\leq\kappa< 10$.
For the Grover walk on a spidernet $S(\kappa,\kappa+2,\kappa-1)$ with 
initial state $\bs{\psi}_0^+$ it holds that
\[
P(X_n=o)
=|\langle\bs{\psi}_0^+,U^n\bs{\psi}_0^+\rangle|^2
\sim 
\left(\frac{10-\kappa}{12}\right)^2\cos^2 n\tilde\theta,
\quad n\rightarrow\infty,
\]
where $\cos\tilde\theta=-1/3$, $0\le \tilde\theta\le \pi$.
Moreover, 
\begin{equation}
\overline{q}^{(\infty)}(o)
=\frac12 \left(\frac{10-\kappa}{12}\right)^2,
\end{equation}
which means that the Grover walk under consideration exhibits initial point localization.
(An example for $\kappa=4$ is shown in Fig.~\ref{fig:one}.)
\end{corollary}

\begin{corollary}\label{3thm:estimate kappa ge10}
Let $\kappa\geq 10$.
For the Grover walk on a spidernet $S(\kappa,\kappa+2,\kappa-1)$ with 
an initial state $\bs{\psi}_0^+$ it holds that
\[
\lim_{n\rightarrow\infty}P(X_n=o)=0,
\quad{hence}\quad
\overline{q}^{(\infty)}(o)=0.
\]
\end{corollary}

\begin{corollary}\label{tree}
For the Grover walk $U$ on a spidernet $S(a,b,b-1)$ with an initial state $\bs{\psi}^+_0$
we have
\[
\lim_{n\rightarrow\infty}P(X_n=o)=0,
\quad{hence}\quad
\overline{q}^{(\infty)}(o)=0.
\]
\end{corollary}

From Corollaries \ref{3cor:k<10}--\ref{tree} we see 
that the localization occurs when the ``density" of large cycles is high.
Further study in this direction is now in progress.

Corollary \ref{tree} follows directly from Theorem \ref{mainthm:probability amplitude}
as a homogeneous tree is a special case of spidernets.
While, quantum walks on a tree have been studied from various aspects and
the result in Corollary \ref{tree} is already known \cite{CHKS}.
Note also that localization may occur for the Grover walk on a tree with 
a non-isotropic initial state.

In relation to Theorem \ref{3thm:localization for spidernet} we have the following

\begin{theorem}\label{exponential localization}
Consider a spidernet $S(a,b,c)$ with $b>c+\sqrt{c}$.
Then for the Grover walk $U$ with an initial state $\bs{\psi}_0^+$ it holds that
\[
\liminf_{N\rightarrow\infty}\frac{1}{N}\sum_{n=0}^{N-1}P(X_n\in V_l)
\ge \frac{b}{2c}\left\{\frac{(b-c)^2-c}{(b-c)(b-c+1)}\right\}^2
    \left\{\frac{c}{(b-c)^2}\right\}^l,
\quad l\ge1.
\]
If the spidernet $S(a,b,c)$ is rotationally symmetric around $o$, we have
\[
\liminf_{N\rightarrow\infty}\frac{1}{N}\sum_{n=0}^{N-1}P(X_n=u)
\ge \frac{b}{2a}\left\{\frac{(b-c)^2-c}{(b-c)(b-c+1)}\right\}^2
    \left\{\frac{1}{(b-c)^2}\right\}^{\partial(u,o)}\,,
\]
for all $u\in V$, $u\neq o$.
Namely, the Grover walk under consideration exhibits exponential localization.
\end{theorem}

Specializing the parameters in Theorem \ref{exponential localization}, we
obtain the following result with no difficulty.

\begin{corollary}\label{local}
For $2\leq\kappa< 10$ the Grover walk on a spidernet $S(\kappa,\kappa+2,\kappa-1)$ with 
an initial state $\bs{\psi}_0^+$ it holds that
\[
\liminf_{N\rightarrow\infty}\frac{1}{N}\sum_{n=0}^{N-1}P(X_n\in V_l)
\ge \frac{\kappa+2}{2(\kappa-1)}\left(\frac{10-\kappa}{12}\right)^2 
 \left(\frac{\kappa-1}{9}\right)^l,
\quad l\ge1.
\]
If the spidernet $S(\kappa,\kappa+2,\kappa-1)$ is rotationally symmetric
around $o$, we have
\[
\liminf_{N\rightarrow\infty}\frac{1}{N}\sum_{n=0}^{N-1}P(X_n=u)
\ge \frac{\kappa+2}{2\kappa}\left(\frac{10-\kappa}{12}\right)^2 
 \left(\frac{1}{9}\right)^{\partial(0,u)},
\quad u\in V, \,\, u\neq o.
\]
Namely, the Grover walk under consideration exhibits exponential localization.
\end{corollary}

\begin{remark}
By changing variable as $\lambda=\cos\theta$,
the right-hand side of \eqref{3eqn:main expression of probability amplitude} becomes an 
integral over $[0,\pi]$. Then using symmetric extension we can write
\[
\int_{-1}^{1} \cos n\theta \,\mu(d\lambda)
=\int_{-\pi}^{\pi} \cos n\theta \,\nu(d\theta)
\]
with a suitable probability distribution $\nu$ on $[-\pi,\pi]$
such that $\nu(-d\theta)=\nu(d\theta)$,
where no point mass at $\pm\pi$.
Thus, we have an alternative expression for the transition amplitude:
\[
\langle \bs{\psi}_0^+, U^n \bs{\psi}_0^+\rangle
=\int_{-\pi}^{\pi} e^{in\theta} \,\nu(d\theta),
\qquad n=0,\pm1,\pm2,\dots,
\]
which is directly related to the spectral decomposition of the unitary
operator $U$.
\end{remark}

\begin{remark}
Let $\{X_n\,;n=0,1,2,\dots\}$ be the isotropic random walk on $S(a,b,c)$ with
transition matrix $T$. It then follows from the well-established general theory that
\begin{equation}\label{3eqn:RW}
P(X_n=o|X_0=o)
=\langle\bs{\delta}_o,T^n\bs{\delta}_o\rangle
=\int_{-1}^{1} \lambda^n \,\mu(d\lambda),
\quad n=0,1,2,\dots,
\end{equation}
where $\mu$ is the same probability distribution as in Theorem \ref{mainthm:probability amplitude},
see also \cite{Obata2012} for relevant discussion along quantum probability.
We see that \eqref{3eqn:RW} makes a good contrast to the transition amplitude 
\eqref{3eqn:main expression of probability amplitude}.
\end{remark}

\section{One-dimensional reduction}
\label{sec:One-dimensional reduction}

\subsection{$(p,q)$-Quantum walk on $\mathbb{Z}_+$}
\label{subsec:4.1}

Let $U=SC$ be the Grover walk on the spidernet $G=S(a,b,c)$.
Define orthonormal vectors in $\mathcal{H}=\ell^2(A(G))$ by
\begin{alignat}{2}
\bs{\psi}^+_n 
&= \frac{1}{\sqrt{ac^n}}
\sum_{u\in V_n}\sum_{\substack{v\in V_{n+1} \\ v\sim u}} \bs{\delta}_{(u,v)}\,,
&\quad &n\ge0, 
\label{3eqn:def of psi+} \\
\bs{\psi}^\circ_n 
&= \frac{1}{\sqrt{a(b-c-1)c^{n-1}}}
\sum_{u\in V_n}\sum_{\substack{v\in V_n \\ v\sim u}} \bs{\delta}_{(u,v)}\,,
&\quad &n\ge1, 
\label{3eqn:def of psio} \\
\bs{\psi}^-_n 
&= \frac{1}{\sqrt{ac^{n-1}}}
\sum_{u\in V_n}\sum_{\substack{v\in V_{n-1} \\ v\sim u}} \bs{\delta}_{(u,v)}\,,
&\quad &n\ge1.
\label{3eqn:def of psi-}
\end{alignat}
We keep the same notations as in \eqref{3eqn:p and q from spidernet}:
\begin{equation}\label{4eqn:pqr}
p=\frac{c}{b}\,,
\qquad
q=\frac{1}{b}\,,
\qquad
r=\frac{b-c-1}{b}\,,
\end{equation}
verifying that
\[
p>0,
\quad q>0, 
\quad r=1-p-q\ge0.
\]

\begin{lemma}\label{lem:actions of C on psi}
It holds that
\begin{align}
C\bs{\psi}^+_n
&=\begin{cases}
\bs{\psi}^+_0\,, & n=0, \\
(2p-1)\bs{\psi}^+_n+2\sqrt{pr}\,\bs{\psi}^\circ_n+2\sqrt{pq}\,\bs{\psi}^-_n\,, & n\ge1,
\end{cases}
\label{eqn:def of psi +n} \\
C\bs{\psi}^\circ_n
&=2\sqrt{pr}\,\bs{\psi}^+_n+(2r-1)\bs{\psi}^\circ_n+2\sqrt{qr}\,\bs{\psi}^-_n\,, 
\,\,\,\,\quad n\ge1,
\label{eqn:def of psi on} \\
C\bs{\psi}^-_n
&=2\sqrt{pq}\,\bs{\psi}^+_n +2\sqrt{qr}\, \bs{\psi}^\circ_n +(2q-1)\bs{\psi}^-_n\,,
\quad n\ge1.
\label{eqn:def of psi -n}
\end{align}
\end{lemma}

\noindent{\it Proof.} 
By definition we have
\begin{align*}
C\bs{\delta}_{(x,y)}
&=\sum_{w\sim x}(H^{(x)})_{yw}\bs{\delta}_{(x,w)}\,,
\qquad (x,y)\in A(G), \\
(H^{(x)})_{yw}
&=
\begin{cases}
\dfrac{2}{a}-\delta_{yw}\,,& x=o, \\[8pt]
\dfrac{2}{b}-\delta_{yw}\,,& \textit{otherwise}.
\end{cases}
\end{align*}
We first show \eqref{eqn:def of psi +n} for $n=0$.
Suppose $(o,y)\in A(G)$.
Then, 
\begin{align*}
C\bs{\delta}_{(o,y)}
&=\sum_{w\sim o}(H^{(o)})_{yw}\bs{\delta}_{(o,w)} \\
&=\sum_{w\sim o}\left(\dfrac{2}{a}-\delta_{yw}\right)\bs{\delta}_{(o,w)} \\
&=\frac{2}{a}\sum_{w\sim o}\bs{\delta}_{(o,w)}-\bs{\delta}_{(o,y)}\,.
\end{align*}
Taking the summation over $y\sim o$, we obtain
\[
\sum_{y\sim o}C\bs{\delta}_{(o,y)}
=2\sum_{w\sim o}\bs{\delta}_{(o,w)}-\sum_{y\sim o}\bs{\delta}_{(o,y)}
=\sum_{w\sim o}\bs{\delta}_{(o,w)}\,,
\]
from which the desired relation follows by
dividing both sides by $\sqrt{a}$.

We next prove \eqref{eqn:def of psi +n} for $n\ge1$.
Suppose $x\in V_n$ with $n\ge1$.
Then by definition,
\begin{align*}
\sum_{\substack{y\in V_{n+1} \\ y\sim x}} C\bs{\delta}_{(x,y)}
&=\sum_{\substack{y\in V_{n+1} \\ y\sim x}} \sum_{w\sim x}\left(\dfrac{2}{b}-\delta_{yw}\right)\bs{\delta}_{(x,w)} \\
&=\frac{2c}{b} \sum_{w\sim x}\bs{\delta}_{(x,w)} 
 - \sum_{\substack{y\in V_{n+1} \\ y\sim x}} \bs{\delta}_{(x,y)}\,,
\end{align*}
where $|\{y\in V_{n+1}\,;\, y\sim x\}|=c$ is taken into account.
Taking the summation over $x\in V_n$, we obtain
\begin{align*}
\sum_{x\in V_n}\sum_{\substack{y\in V_{n+1} \\ y\sim x}} C\bs{\delta}_{(x,y)}
&=\frac{2c}{b} \sum_{x\in V_n}\sum_{w\sim x}\bs{\delta}_{(x,w)} 
 - \sum_{x\in V_n}\sum_{\substack{y\in V_{n+1} \\ y\sim x}} \bs{\delta}_{(x,y)} \\
&=\frac{2c}{b}\Big(\sqrt{ac^n}\, \bs{\psi}^+_n
 + \sqrt{a(b-c-1)c^{n-1}}\,\bs{\psi}^\circ_n \\
&\qquad\qquad  +\sqrt{ac^{n-1}}\,\bs{\psi}^-_n\Big) 
 - \sqrt{ac^n}\,\bs{\psi}^+_n(x)
\end{align*}
and then, dividing both sides by $\sqrt{ac^n}$, we come to
\[
C\bs{\psi}^+_n
=\left(\frac{2c}{b}-1\right)\bs{\psi}^+_n
 +\frac{2\sqrt{c(b-c-1)}}{b}\,\bs{\psi}^\circ_n
 +\frac{2\sqrt{c}}{b}\,\bs{\psi}^-_n\,,
\]
which shows \eqref{eqn:def of psi +n}.
The rest of the relations is proved in a similar manner.
\begin{flushright} $\square$ \end{flushright}

\begin{lemma}\label{docomo}
It holds that
\begin{alignat}{2}
S\bs{\psi}^+_n &=\bs{\psi}^-_{n+1}\,, &\quad &n\ge0,
\label{eqn:def of S psi +n} \\
S\bs{\psi}^\circ_n &=\bs{\psi}^\circ_n\,, &\quad &n\ge1, 
\label{eqn:def of S psi on} \\
S\bs{\psi}^-_n &= \bs{\psi}^+_{n-1}\,, &\quad &n\ge1.
\label{eqn:def of S psi -n}
\end{alignat}
\end{lemma}

\noindent{\it Proof.}
By Straightforward calculation similar to the proof of Lemma \ref{lem:actions of C on psi}.
\begin{flushright}$\square$\end{flushright}

\noindent It is convenient to study the actions of $C$ and $S$
described in Lemmas \ref{lem:actions of C on psi} and \ref{docomo}
in a slightly more general context.
We consider the Hilbert space $\mathcal{H}(\mathbb{Z}_+)$ of the form:
\[
\mathcal{H}(\mathbb{Z}_+)
=\mathbb{C}\bs{\psi}^+_0\oplus \sum_{n=1}^\infty\oplus 
 (\mathbb{C}\bs{\psi}^+_n\oplus \mathbb{C}\bs{\psi}^\circ_n\oplus \mathbb{C}\bs{\psi}^-_n),
\]
where $\bs{\psi}^+_0, \bs{\psi}^+_1,\bs{\psi}^\circ_1, \bs{\psi}^-_1,\dots$ form an 
orthonormal basis of $\mathcal{H}(\mathbb{Z}_+)$.
Let $p,q,r$ be constant numbers satisfying
\[
p>0, \quad q>0, \quad r=1-p-q\ge0.
\]
We then define the coin operator $C$ and the shift operator $S$ on $\mathcal{H}(\mathbb{Z}_+)$
by \eqref{eqn:def of psi +n}--\eqref{eqn:def of psi -n} and by \eqref{eqn:def of S psi +n}--\eqref{eqn:def of S psi -n},
respectively. 
It is easily seen that both $C$ and $S$ are unitary operators.
Hence $U=SC$ is also a unitary operator on $\mathcal{H}(\mathbb{Z}_+)$,
which is called the \textit{$(p,q)$-quantum walk} on $\mathbb{Z}_+$.

Thus the Grover walk on a spidernet $G=S(a,b,c)$ restricted to
the closed subspace spanned by
$\{\bs{\psi}^+_n,\,n\ge0\}\cup
 \{\bs{\psi}^\circ_n\,;\, n\ge1\}\cup
 \{\bs{\psi}^-_n\,;\,n\ge1\}$ is a $(p,q)$-quantum walk on $\mathbb{Z}_+$,
where $p,q$ are given by \eqref{4eqn:pqr}.

We define orthonormal vectors in $\mathcal{H}(\mathbb{Z}_+)$ by
\begin{align*}
\bs{\Psi}_0 &=\bs{\psi}^+_0, \\
\bs{\Psi}_n &=\sqrt{p}\,\bs{\psi}^+_n+\sqrt{r}\,\bs{\psi}^\circ_n+\sqrt{q}\,\bs{\psi}^-_n\,, \quad n\ge1,
\end{align*}
and set
\[
\Gamma(\mathbb{Z}_+)=\sum_{n=0}^\infty \oplus \mathbb{C} \bs{\Psi}_n\,.
\]
Then $\Gamma(\mathbb{Z}_+)\subset \mathcal{H}(\mathbb{Z}_+)$ is a closed subspace.
Let $\Pi:\mathcal{H}(\mathbb{Z}_+)\rightarrow\Gamma(\mathbb{Z}_+)$ denote the orthogonal projection.

\begin{lemma}\label{3lem:C=2pi-I on Z}
It holds that
\[
C=C^*=2\Pi-I.
\]
In particular, $C$ is the reflection with respect to $\Gamma(\mathbb{Z}_+)$ and
acts on $\Gamma(\mathbb{Z}_+)$ as the identity.
\end{lemma}

\noindent{\it Proof.}
Straightforward by definition.
\begin{flushright}$\square$\end{flushright}

\subsection{$(p,q)$-Quantum walk on a path of finite length}

Let $U$ be a $(p,q)$-quantum walk on $\mathbb{Z}_+$ as in the previous section.
We will introduce a $(p,q)$-quantum walk on the path of length $N\ge2$,
obtained from the $(p,q)$-quantum walk on $\mathbb{Z}_+$ by cutoff.

For $N\ge2$ we define a Hilbert space:
\[
\mathcal{H}(N)
=\mathbb{C}\bs{\psi}^+_0\oplus \sum_{n=1}^{N-1}\oplus 
 (\mathbb{C}\bs{\psi}^+_n\oplus \mathbb{C}\bs{\psi}^\circ_n\oplus \mathbb{C}\bs{\psi}^-_n)
 \oplus \mathbb{C}\bs{\psi}^-_N
\]
and unitary operators $C=C_N$ and $S=S_N$ respectively 
as in \eqref{eqn:def of psi +n}--\eqref{eqn:def of psi -n}
and in \eqref{eqn:def of S psi +n}--\eqref{eqn:def of S psi -n}, except
\begin{equation}\label{3eqn:exception C}
C\bs{\psi}^-_N=\bs{\psi}^-_N.
\end{equation}
Then we obtain a unitary operator $U=U_N=S_NC_N$ on $\mathcal{H}(N)$,
which is called the \textit{$(p,q)$-quantum walk} on the path of length $N$.
Both endpoints play as reflection barriers in analogy of random walks.
From now on we omit the suffix $N$ whenever there is no danger of confusion.

In view of \eqref{eqn:def of psi +n}--\eqref{eqn:def of psi -n}, \eqref{3eqn:exception C}
and \eqref{eqn:def of S psi +n}--\eqref{eqn:def of S psi -n}
the explicit actions of $U$ on $\bs{\psi}^\epsilon_j$ are easily written down as follows:
\begin{align}
U\bs{\psi}^+_j
&=\begin{cases}
\bs{\psi}^-_1\,, & j=0, \\
(2p-1)\bs{\psi}^-_{j+1}+2\sqrt{pr}\,\bs{\psi}^\circ_j+2\sqrt{pq}\,\bs{\psi}^+_{j-1}\,, & 1\le j\le N-1,
\end{cases}
\label{eqn:action of U psi +n} \\
U\bs{\psi}^\circ_j
&=2\sqrt{pr}\,\bs{\psi}^-_{j+1}+(2r-1)\bs{\psi}^\circ_j+2\sqrt{qr}\,\bs{\psi}^+_{j-1}\,, 
\,\,\,\,\quad 1\le j\le N-1,
\label{eqn:action of U psi on} \\
U\bs{\psi}^-_j
&=\begin{cases}
2\sqrt{pq}\,\bs{\psi}^-_{j+1} +2\sqrt{qr}\, \bs{\psi}^\circ_j +(2q-1)\bs{\psi}^+_{j-1}\,, & 1\le j\le N-1, \\
\bs{\psi}^+_{N-1}, & j=N,
\end{cases}
\label{eqn:action of U psi -n}
\end{align}

The goal of this subsection is to determine the spectra (eigenvalues) of $U$.
We start with the following result.

\begin{lemma}\label{3lem:Trace of U}
$\mathrm{Tr\,}U=(2r-1)(N-1)$.
\end{lemma}

\noindent{\it Proof.}
We see from \eqref{eqn:action of U psi +n}--\eqref{eqn:action of U psi -n} that
\[
\mathrm{Tr\,}U=\sum_{\epsilon,n}\langle \bs{\psi}^\epsilon_n, U \bs{\psi}^\epsilon_n\rangle
=\sum_{j=1}^{N-1} \langle \bs{\psi}^\circ_j, U \bs{\psi}^\circ_j\rangle
=(2r-1)(N-1)
\]
as desired.
\begin{flushright}$\square$\end{flushright}

\noindent Define orthonormal vectors in $\mathcal{H}(N)$ by
\begin{align*}
&\bs{\Psi}_0 =\bs{\psi}^+_0, \\
&\bs{\Psi}_j =\sqrt{p}\,\bs{\psi}^+_j+\sqrt{r}\,\bs{\psi}^\circ_j+\sqrt{q}\,\bs{\psi}^-_j\,, \quad 1\le j\le N-1, \\
&\bs{\Psi}_N =\bs{\psi}^-_N
\end{align*}
and set
\[
\Gamma(N)=\sum_{j=0}^N \oplus\mathbb{C} \bs{\Psi}_j\,.
\]
Then $\Gamma(N)$ is a closed subspace of $\mathcal{H}(N)$
and let $\Pi=\Pi_N:\mathcal{H}(N)\rightarrow\Gamma(N)$ denote the orthogonal projection.
The assertion of Lemma \ref{3lem:C=2pi-I on Z} remains true, i.e., 
it holds that 
\[
C=C^*=2\Pi-I.
\]

For the $(p,q)$-quantum walk $U=U_N$ we consider $\Pi U \Pi$ as an operator on $\Gamma(N)$, 
which is denoted by $T=T_N$.
Thus,
\begin{equation}\label{4eqn:T and Pi}
T=\Pi U \Pi \!\!\restriction_{\Gamma(N)}
=\Pi SC \Pi \!\!\restriction_{\Gamma(N)}
=\Pi S \!\!\restriction_{\Gamma(N)}\,.
\end{equation}
Moreover, by direct calculation we obtain its matrix expression 
with respect to the orthonormal basis $\{\bs{\Psi}_j\,;\,0\le j\le N\}$ as follows:
\[
T=T_N=
\begin{bmatrix}
0        & \sqrt{q} &                                \\
\sqrt{q} & r        & \sqrt{pq} &                    \\  
         & \sqrt{pq}& r         & \sqrt{pq} &        \\  
         &          & \ddots    & \ddots    & \ddots \\  
& & &  \sqrt{pq} & r        &\sqrt{pq} \\  
& & & & \sqrt{pq} & r        &\sqrt{p} \\  
& & & &           & \sqrt{p} & 0 \\  
\end{bmatrix}.
\]
For example,
\begin{align*}
&T \bs{\Psi}_0 =\sqrt{q}\, \bs{\Psi}_1\,, \\
&T \bs{\Psi}_1 =\sqrt{pq}\, \bs{\Psi}_2+ r \bs{\Psi}_1 + \sqrt{q}\,\bs{\Psi}_0\,,
\quad \text{etc.}
\end{align*}

\begin{lemma}
Every eigenvalue of $T$ is simple.
Moreover, $\mathrm{Spec}(T)\subset [-1,1]$ and $1\in \mathrm{Spec}(T)$.
\end{lemma}

\noindent{\it Proof.}
That every eigenvalue of $T$ is simple follows from general theory of Jacobi matrices
(see e.g., \cite{HO,Deift}).
Since $T=\Pi S\!\!\restriction_{\Gamma(N)}$ by \eqref{4eqn:T and Pi}, the operator norm of $T$ is bounded by one.
Hence every eigenvalue of $T$ lies in $[-1,1]$.
Finally, it is easily verified by expansion that $\det(T-I)=0$.
\begin{flushright}$\square$\end{flushright}

\begin{lemma}\label{3lem:3.6}
{\upshape (1)} If $r>0$, there exists no non-zero $v\in\Gamma(N)$ such that $Sv=-v$.

{\upshape (2)} If $r=0$, there exists non-zero $v\in\Gamma(N)$ such that $Sv=-v$.
Moreover, such a non-zero vector $v$ is determined uniquely up to a constant factor.

\end{lemma}

\noindent{\it Proof.}
Every $v\in \Gamma(N)$ is in the form:
\begin{align*}
v
&=\sum_{j=0}^N \gamma_j \bs{\Psi}_j \\
&=\gamma_0 \bs{\psi}_0^+
  +\sum_{j=1}^{N-1} \gamma_j (\sqrt{p}\,\bs{\psi}_j^+ +\sqrt{r}\, \bs{\psi}_j^\circ + \sqrt{q}\, \bs{\psi}_j^-)
  +\gamma_N \bs{\psi}_N^-,
\end{align*}
where $\gamma_0,\dots,\gamma_N$ are constant numbers.
Then the equation $Sv=-v$ is equivalent to the one for these constant numbers,
which is obtained by direct calculation:
\begin{align}
&\gamma_1=-\frac{1}{\sqrt{q}}\,\gamma_0\,,
\quad
\gamma_N=-\sqrt{p}\,\gamma_{N-1}\,, 
\label{4eqn:in proof (000)} \\
&\gamma_j=-\sqrt{\frac{p}{q}}\,\gamma_{j-1}\,,
\quad 2\le j\le N-1,
\label{4eqn:in proof (001)} \\
&\gamma_j\sqrt{r}=-\gamma_j \sqrt{r}\,,
\quad 1\le j\le N-1.
\label{4eqn:in proof (002)} 
\end{align}
If $r>0$, it follows from \eqref{4eqn:in proof (002)} that $\gamma_1=\dots=\gamma_{N-1}=0$.
Then in view of \eqref{4eqn:in proof (000)} we also have $\gamma_0=\gamma_N=0$,
which implies $v=0$.
If $r=0$, the recurrence relations \eqref{4eqn:in proof (000)} and \eqref{4eqn:in proof (001)} determine
the sequence $\gamma_0,\gamma_1,\dots,\gamma_N$ uniquely by the initial value $\gamma_0$.
Hence $\dim\{v\in\Gamma(N)\,;\, Sv=-v\}=1$ as desired.
\begin{flushright}$\square$\end{flushright}

\begin{lemma}\label{3lem:3.7}
If $Tv=\pm v$ for $v\in\Gamma(N)$, then $Sv=Uv=\pm v$.
\end{lemma}

\noindent{\it Proof.}
It is sufficient to consider the case of $v\neq0$.
Suppose that $Tv=\pm v$ for $v\in\Gamma(N)$.
From $T=\Pi S\!\!\restriction_{\Gamma(N)}$ and $\Pi v=v$ we obtain $\Pi(Sv\mp v)=0$.
Hence $\langle Sv \mp v,v\rangle=0$, which implies that
\[
\langle Sv,v \rangle
=\pm \langle v,v \rangle.
\]
Since $S$ is unitary, the above relation implies the Schwartz equality
and $Sv=\alpha v$ with some constant $\alpha\in\mathbb{C}$.
It follows by applying $\Pi$ that $\alpha=\pm1$ and $Sv=\pm v$.
Finally, since $U=SC$ by definition and $C$ acts on $\Gamma(N)$ as the identity,
we have $Uv=SCv=Sv$.
\begin{flushright}$\square$\end{flushright}

\begin{lemma}
We have $-1 \not\in\mathrm{Spec}(T)$ for $r>0$,
and $-1 \in \mathrm{Spec}(T)$ for $r=0$.
\end{lemma}

\noindent{\it Proof.}
Suppose that $r>0$ and $Tv=-v$ for $v\in\Gamma(N)$.
We see from Lemma \ref{3lem:3.7} $Sv=-v$,
then applying Lemma \ref{3lem:3.6} we come to $v=0$.
This means that $-1$ is not an eigenvalue of $T$.

We next suppose that $r=0$. 
By Lemma \ref{3lem:3.6} there exists a non-zero vector $v\in\Gamma(N)$ such that $Sv=-v$.
Then $Tv=\Pi Sv=-v$, which means that $-1$ is an eigenvalue of $T$.
\begin{flushright}$\square$\end{flushright}

\noindent Thus, the eigenvalues of $T$ are arranged in such a way that
\begin{gather}
\lambda_0=1=\cos\theta_0,
\quad
\lambda_1=\cos\theta_1\,,
\quad
\lambda_2=\cos\theta_2\,,
\quad \dots,
\quad
\lambda_N=\cos\theta_N\,, 
\label{3eqn:arrangement of lambda} \\
0=\theta_0<\theta_1<\theta_2<\dots<\theta_N\le \pi,
\nonumber
\end{gather}
where $\theta_N<\pi$ for $r>0$ and $\theta_N=\pi$ for $r=0$.
For each eigenvalue $\lambda_j$ we take a normalized eigenvector $\bs{\Omega}_j\in\Gamma(N)$,
i.e.,
\[
T\bs{\Omega}_j=\lambda_j \bs{\Omega}_j\,,
\quad
\|\bs{\Omega}_j\|=1.
\]
Then we have the orthogonal decomposition of $\Gamma(N)$ in two ways:
\[
\Gamma(N)
=\sum_{j=0}^N \oplus \mathbb{C}\bs{\Psi}_j
=\sum_{j=0}^N \oplus \mathbb{C}\bs{\Omega}_j.
\]
We next study the subspace
\begin{equation}\label{4eqn:def of L(N)}
\mathcal{L}(N)=\Gamma(N)+S\Gamma(N),
\end{equation}
which is invariant under the actions of $S$ and $U$.
In fact, for $\bs{\phi}, \bs{\psi}\in \Gamma(N)$ we have
\begin{align*}
U(\bs{\phi}+S\bs{\psi})
&=SC\bs{\phi}+SCS\bs{\psi}
=S\bs{\phi}+S(2\Pi-I)S\bs{\psi} \\
&=S\bs{\phi}+2S\Pi S\bs{\psi}-S^2\bs{\psi}
=S(\bs{\phi}+2\Pi S\bs{\psi})-\bs{\psi},
\end{align*}
which shows that $\mathcal{L}(N)=\Gamma(N)+S\Gamma(N)$ is invariant under $U$.

\begin{lemma}\label{3lem:orthogonality}
{\upshape (1)} If $r>0$, then the vectors $\bs{\Omega}_0,\bs{\Omega}_1,\dots, \bs{\Omega}_N,
S\bs{\Omega}_1,\dots, S\bs{\Omega}_N$ are linearly independent.
Moreover,
\begin{equation}\label{3eqn:inner product 3.9}
\langle \bs{\Omega}_j, S\bs{\Omega}_k\rangle=\lambda_k\delta_{jk}\,,
\quad 0\le j\le N, 
\quad 1\le k\le N.
\end{equation}

{\upshape (2)} If $r=0$, then $S\bs{\Omega}_N=-\bs{\Omega}_N$ and 
$\bs{\Omega}_0,\bs{\Omega}_1,\dots, \bs{\Omega}_N$,
$S\bs{\Omega}_1,\dots, S\bs{\Omega}_{N-1}$ are linearly independent.
Moreover, \eqref{3eqn:inner product 3.9} remains valid where
$0\le j\le N$ and $1\le k\le N-1$.
\end{lemma}

\noindent{\it Proof.}
(1) Suppose that
\begin{equation}\label{3eqn:in proof 3.9 (1)}
\alpha_0 \bs{\Omega}_0 
+\sum_{j=1}^N \alpha_j \bs{\Omega}_j
+\sum_{j=1}^N \beta_j S\bs{\Omega}_j=0.
\end{equation}
Taking $\Pi S=T$ in mind, we apply $\Pi$ to both sides to obtain
\begin{equation}\label{3eqn:in proof 3.9 (2)}
\alpha_0 \bs{\Omega}_0 +\sum_{j=1}^N (\alpha_j+\beta_j\lambda_j) \bs{\Omega}_j=0.
\end{equation}
Similarly, applying $\Pi S$ to both sides of \eqref{3eqn:in proof 3.9 (1)}, we obtain
\begin{equation}\label{3eqn:in proof 3.9 (3)}
\alpha_0 \bs{\Omega}_0 +\sum_{j=1}^N (\alpha_j\lambda_j+\beta_j) \bs{\Omega}_j=0,
\end{equation}
where $S^2=I$ and $S\bs{\Omega}_0=\bs{\Omega}_0$ from Lemma \ref{3lem:3.7} are taken into account.
It then follows from \eqref{3eqn:in proof 3.9 (2)} and \eqref{3eqn:in proof 3.9 (3)} that
\[
\alpha_0=0, \quad \alpha_j+\beta_j\lambda_j=\alpha_j\lambda_j+\beta_j=0,
\quad 1\le j\le N.
\]
Since $\lambda_j\neq \pm1$ for $1\le j\le N$, we see that $\alpha_j=\beta_j=0$ for all $j$.
The inner product \eqref{3eqn:inner product 3.9} is computed as follows:
\[
\langle \bs{\Omega}_j, S\bs{\Omega}_k\rangle
=\langle \bs{\Omega}_j, \Pi S\bs{\Omega}_k\rangle
 =\langle \bs{\Omega}_j, T\bs{\Omega}_k\rangle
=\lambda_k \langle \bs{\Omega}_j, \bs{\Omega}_k\rangle
=\lambda_k\delta_{kj}\,.
\]

(2) is proved similarly by
using Lemma \ref{3lem:3.7} and $\lambda_N=-1$. 
\begin{flushright}$\square$\end{flushright}

\begin{lemma}\label{4lem:U on omegas}
{\upshape (1)} If $r>0$ we have
\begin{align*}
&U \bs{\Omega}_0=\bs{\Omega}_0, \\
&U \bs{\Omega}_j=S\bs{\Omega}_j, \quad
U S\bs{\Omega}_j=-\bs{\Omega}_j+2\lambda_j S\bs{\Omega}_j, 
\quad 1\le j\le N.
\end{align*}

{\upshape (2)} If $r=0$, the above relations hold except $U \bs{\Omega}_N=-\bs{\Omega}_N$.
\end{lemma}

\noindent{\it Proof.}
Since the proofs are similar we prove only (1).
We first observe that
\[
U\bs{\Omega}_j
=SC\bs{\Omega}_j
=S(2\Pi-I)\bs{\Omega}_j
=2S\Pi\bs{\Omega}_j -S\bs{\Omega}_j
=S\bs{\Omega}_j\,,
\quad 0\le j\le N.
\]
For $j=0$ we have $T\bs{\Omega}_0=\bs{\Omega}_0$ so 
$S\bs{\Omega}_0=\bs{\Omega}_0$ by Lemma \ref{3lem:3.7}.
Hence
\[
U\bs{\Omega}_0=SC\bs{\Omega}_0=\bs{\Omega}_0\,.
\]
We next calculate $US\bs{\Omega}_j$ for $1\le j\le N$. 
Using $C=2\Pi-I$ and $S^2=I$ we obtain
\begin{align*}
US\bs{\Omega}_j
&=SCS\bs{\Omega}_j
=S(2\Pi-I)S\bs{\Omega}_j \\
&=2ST\bs{\Omega}_j-\bs{\Omega}_j
=2\lambda_j S\bs{\Omega}_j-\bs{\Omega}_j\,.
\end{align*}
\begin{flushright}$\square$\end{flushright}

\noindent Thus, we obtain the orthogonal decomposition of $\mathcal{L}(N)$ defined
in \eqref{4eqn:def of L(N)}:
\begin{alignat}{2}
\mathcal{L}(N)
&=\mathbb{C}\bs{\Omega}_0 \oplus
 \sum_{j=1}^N \oplus (\mathbb{C}\bs{\Omega}_j+\mathbb{C}S\bs{\Omega}_j),
&\qquad &r>0, 
\label{4eqn:decomposition of L(N)}\\
\mathcal{L}(N)
&=\mathbb{C}\bs{\Omega}_0 \oplus
 \sum_{j=1}^{N-1} \oplus (\mathbb{C}\bs{\Omega}_j+\mathbb{C}S\bs{\Omega}_j)
 \oplus \mathbb{C}\bs{\Omega}_N\,,
&\quad &r=0,
\label{4eqn:decomposition of L(N)(r=0)}
\end{alignat}
where each factor is invariant under the action of $U$.

\begin{theorem}\label{3thmspectra of U}
{\upshape (1)} If $r>0$, the eigenvalues of $U$ are
\[
1, 
\quad e^{\pm i\theta_j} \quad (1\le j\le N),
\quad -1,
\]
where $0<\theta_1<\theta_2<\dots<\theta_N<\pi$
are obtained in \eqref{3eqn:arrangement of lambda}.
All the eigenvalues except $-1$ are multiplicity free
and the multiplicity of the eigenvalue $-1$ is $N-2$.

{\upshape (2)} If $r=0$, the eigenvalues of $U$ are
\[
1, 
\quad e^{\pm i\theta_j} \quad (1\le j\le N-1),
\quad -1,
\]
where $0<\theta_1<\theta_2<\dots<\theta_{N-1}<\pi$.
All the eigenvalues except $-1$ are multiplicity free
and the multiplicity of the eigenvalue $-1$ is $N$.
\end{theorem}

\noindent{\it Proof.}
Since the proofs are similar we prove only (1).
The orthogonal decomposition \eqref{4eqn:decomposition of L(N)} gives rise to 
a blockwise diagonalization of $U$.
It is obvious from Lemma \ref{4lem:U on omegas} 
that $U$ restricted to $\mathbb{C}\bs{\Omega}_0$ is the identity operator.
Next suppose that $1\le j\le N$.
We see from Lemma \ref{4lem:U on omegas} that $U$
restricted to $\mathbb{C}\bs{\Omega}_j+\mathbb{C}S\bs{\Omega}_j$ 
admits a matrix representation:
\[
\begin{bmatrix}
0 & -1 \\
1 & 2\lambda_j
\end{bmatrix},
\]
of which the eigenvalues are 
\[
\lambda_j\pm i\sqrt{1-\lambda_j^2}=e^{\pm i\theta_j}.
\]
Denoting by $\mathcal{M}$ the orthogonal complement of $\mathcal{L}(N)$ in $\mathcal{H}(N)$,
we have
\begin{align*}
\mathrm{Tr\,}(U)
&=1+\sum_{j=1}^N2\lambda_j + \mathrm{Tr\,}(U\!\!\restriction_{\mathcal{M}})
=1+2(\mathrm{Tr\,}(T)-1)+\mathrm{Tr\,}(U\!\!\restriction_{\mathcal{M}}) \\
&=2\mathrm{Tr\,}(T)-1+\mathrm{Tr\,}(U\!\!\restriction_{\mathcal{M}})
=2r(N-1)-1+\mathrm{Tr\,}(U\!\!\restriction_{\mathcal{M}}).
\end{align*}
On the other hand, $\mathrm{Tr\,}(U)=(2r-1)(N-1)$ by Lemma \ref{3lem:Trace of U}.
Hence
\[
\mathrm{Tr\,}(U\!\!\restriction_{\mathcal{M}})
=(2r-1)(N-1)-(2r(N-1)-1)
=-(N-2).
\]
Since $\dim \mathcal{M}=(3N-1)-(2N+1)=N-2$,
we see that $U\!\!\restriction_{\mathcal{M}}=-I$.
Therefore $\mathcal{M}$ is the eigenspace of $U$ with eigenvalue $-1$
so that the multiplicity of the eigenvalue $-1$ coincides with $\dim \mathcal{M}=N-2$.
\begin{flushright}$\square$\end{flushright}

\begin{theorem}
{\upshape (1)} Let $r>0$ and set 
\[
\bs{\Omega}_j^{\pm}=\frac{1}{\sqrt2\, \sin\theta_j}\,(\bs{\Omega}_j-e^{\pm i\theta_j}S\bs{\Omega}_j),
\quad 1\le j\le N.
\]
Then $\bs{\Omega}_j^{\pm}\in (\mathbb{C}\bs{\Omega}_j+\mathbb{C}S\bs{\Omega}_j)$, 
$\|\bs{\Omega}_j^{\pm}\|=1$ and
\[
U\bs{\Omega}_j^{\pm}=e^{\pm i\theta_j} \bs{\Omega}_j^{\pm}.
\]
In other words,
$\bs{\Omega}_j^{\pm}$ is a normalized eigenvector of $U$ with eigenvalue $e^{\pm i\theta_j}$.

{\upshape (2)} If $r=0$, the above assertion remains valid for $1\le j\le N-1$.
\end{theorem}

\noindent{\it Proof.}
That $\|\bs{\Omega}_j^{\pm}\|=1$ is verified by using Lemma \ref{3lem:orthogonality}.
That $U\bs{\Omega}_j^{\pm}=e^{\pm i\theta_j} \bs{\Omega}_j^{\pm}$ follows 
from Lemma \ref{4lem:U on omegas}.
\begin{flushright}$\square$\end{flushright}

\section{Proofs of main results}
\label{sec:Proofs of main results}

\subsection{Transition amplitudes of the $(p,q)$-quantum walk on $\mathbb{Z}_+$}

For Theorem \ref{mainthm:probability amplitude} we need to calculate
the transition amplitude:
\begin{equation}\label{5eqn:goal amplitude}
\langle \bs{\psi}_0^+, U^n \bs{\psi}_0^+\rangle
=\langle \bs{\Psi}_0, U^n \bs{\Psi}_0\rangle
\end{equation}
for the $(p,q)$-quantum walk $U$ on $\mathbb{Z}_+$.
A key observation here is that \eqref{5eqn:goal amplitude} is 
calculated after cutoff.
More generally, if $\bs{\phi}, \bs{\psi}\in\mathcal{H}(\mathbb{Z}_+)$ have
finite supports, 
\begin{equation}\label{5eqn:goal amplitude finite}
\langle \bs{\phi}, U^n \bs{\psi}\rangle
=\langle \bs{\phi}, U_N^n \bs{\psi}\rangle_{\mathcal{L}(N)}
\end{equation}
holds for all sufficiently large $N$.
In fact, if $\mathrm{supp\,}\bs{\phi}\subset\{0,1,\dots,l\}$ and
$\mathrm{supp\,} \bs{\psi}\subset\{0,1,\dots,m\}$, then
\eqref{5eqn:goal amplitude finite} holds for $N>\min\{n+l,n+m\}$.
The purpose of this subsection is to derive an integral formula for \eqref{5eqn:goal amplitude finite}.

Now let $N\ge2$ be fixed and start with 
the $(p,q)$-quantum walk $U$ on the path of length $N\ge2$.

\begin{lemma}\label{4lem:4.33}
For $r>0$ it holds that
\begin{align}
\langle \bs{\Psi}_l,\bs{\Omega}_j^{\pm} \rangle 
&= \frac{\mp ie^{\pm i\theta_j}}{\sqrt{2}}\langle \bs{\Psi}_l,\bs{\Omega}_j \rangle, 
\qquad 1\le j\le N,
\label{4eqn:in 4.33 (1)}\\
\langle S\bs{\Psi}_l,\bs{\Omega}_0 \rangle
&=\langle \bs{\Psi}_l,\bs{\Omega}_0 \rangle, 
\label{4eqn:in 4.33 (3)}\\
\langle S\bs{\Psi}_l,\bs{\Omega}_j^{\pm} \rangle 
&= \frac{\mp i}{\sqrt{2}}\langle \bs{\Psi}_l,\bs{\Omega}_j \rangle,
\qquad 1\le j\le N,
\label{4eqn:in 4.33 (2)}
\end{align}
For $r=0$ the above relations remain valid for $1\le j\le N-1$ and
\begin{equation}\label{4eqn:in 4.33 (6)}
\langle S\bs{\Psi}_l,\bs{\Omega}_N \rangle
=-\langle \bs{\Psi}_l,\bs{\Omega}_N \rangle. 
\end{equation}
\end{lemma}

\noindent{\it Proof.}
By definition we have
\begin{equation}\label{4eqn:in proof 4.2 (1)}
\langle\bs{\Psi}_l,\bs{\Omega}_j^{\pm}\rangle
=\frac{1}{\sqrt2\,\sin\theta_j}\,\left(
 \langle\bs{\Psi}_l,\bs{\Omega}_j\rangle
 -e^{\pm i\theta_j}\langle\bs{\Psi}_l,S\bs{\Omega}_j\rangle
 \right).
\end{equation}
Since $\Pi\bs{\Psi}_l=\bs{\Psi}_l$ we have
\[
\langle\bs{\Psi}_l,S\bs{\Omega}_j\rangle
= \langle\bs{\Psi}_l,\Pi S\bs{\Omega}_j\rangle
= \langle\bs{\Psi}_l,T\bs{\Omega}_j\rangle
= \lambda_j \langle\bs{\Psi}_l,\bs{\Omega}_j\rangle.
\]
Then \eqref{4eqn:in proof 4.2 (1)} becomes
\begin{equation}\label{4eqn:in proof 4.2 (2)}
\langle\bs{\Psi}_l,\bs{\Omega}_j^{\pm}\rangle
=\frac{1}{\sqrt2\,\sin\theta_j}\,(1-e^{\pm i\theta_j}\lambda_j) \langle\bs{\Psi}_l,\bs{\Omega}_j\rangle.
\end{equation}
We see easily from  $\cos\theta_j=\lambda_j$ that
\[
1-e^{\pm i\theta_j}\lambda_j=\mp i (\sin \theta_j) e^{\pm i\theta_j}.
\]
Inserting the above relation into \eqref{4eqn:in proof 4.2 (2)}, we obtain \eqref{4eqn:in 4.33 (1)}. 

We next show \eqref{4eqn:in 4.33 (2)}.
In view of the definition of $\bs{\Omega}_j^{\pm}$ and using $S^2=I$ we have
\begin{align*}
\langle S\bs{\Psi}_l,\bs{\Omega}_j^{\pm}\rangle
&=\frac{1}{\sqrt2\,\sin\theta_j}\,\left(
 \langle S\bs{\Psi}_l,\bs{\Omega}_j\rangle
 -e^{\pm i\theta_j}\langle S\bs{\Psi}_l,S\bs{\Omega}_j\rangle
 \right) \\
&=\frac{1}{\sqrt2\,\sin\theta_j}\,\left(
 \langle \bs{\Psi}_l, S\bs{\Omega}_j\rangle
 -e^{\pm i\theta_j}\langle \bs{\Psi}_l,\bs{\Omega}_j\rangle
 \right). 
\end{align*}
Then, applying a similar consideration as above, we obtain
\eqref{4eqn:in 4.33 (2)} with no difficulty.

Finally, since $U\bs{\Omega}_0=S\bs{\Omega}_0=\bs{\Omega}_0$, we have 
\[
\langle S\bs{\Psi}_l,\bs{\Omega}_0 \rangle
=\langle \bs{\Psi}_l, S\bs{\Omega}_0 \rangle
=\langle \bs{\Psi}_l,\bs{\Omega}_0 \rangle, 
\]
which shows \eqref{4eqn:in 4.33 (3)}.
For \eqref{4eqn:in 4.33 (6)} we need only to note that $S\bs{\Omega}_N=-\bs{\Omega}_N$.
\begin{flushright}$\square$\end{flushright}

\begin{lemma}\label{5lem:four inner products}
For $0\le l,m\le N$ and $n=0,\pm1,\pm2,\dots$ it holds that
\begin{align}
\langle \bs{\Psi}_l,U^n\bs{\Psi}_m \rangle
&=\sum_{j=0}^N (\cos n\theta_j)
 \langle \bs{\Psi}_l, \bs{\Omega}_j\rangle
 \langle \bs{\Omega}_j, \bs{\Psi}_m\rangle,  
\label{4eqn:5-2-01} \\
\langle S\bs{\Psi}_l,U^n\bs{\Psi}_m \rangle
&=\langle \bs{\Psi}_l,U^{n-1}\bs{\Psi}_m \rangle, 
\label{4eqn:5-2-02} \\
\langle \bs{\Psi}_l,U^n S\bs{\Psi}_m \rangle
&=\langle \bs{\Psi}_l,U^{n+1}\bs{\Psi}_m \rangle, 
\label{4eqn:5-2-03} \\
\langle S\bs{\Psi}_l, U^n S\bs{\Psi}_m \rangle
&=\langle \bs{\Psi}_l,U^{n}\bs{\Psi}_m \rangle.
\label{4eqn:5-2-04}
\end{align}
\end{lemma}

\noindent{\it Proof.}
Because the proofs are similar, we prove the assertions under $r>0$. 
Since $\bs{\Psi}_l$ and $U^n\bs{\Psi}_m$ are vectors in $\mathcal{L}(N)$, 
the left-hand side of \eqref{4eqn:5-2-01}
is expanded in terms of the orthonormal basis
$\bs{\Omega}_0, \bs{\Omega}_1^\pm,\dots, \bs{\Omega}_N^\pm$ as follows:
\begin{align}
\langle \bs{\Psi}_l,U^n\bs{\Psi}_m \rangle
&=\langle \bs{\Psi}_l,\bs{\Omega}_0\rangle\langle \bs{\Omega}_0, U^n\bs{\Psi}_m \rangle
 +\sum_{j=1}^N \langle \bs{\Psi}_l,\bs{\Omega}_j^+\rangle\langle \bs{\Omega}_j^+, U^n\bs{\Psi}_m \rangle 
\nonumber \\
&\qquad  +\sum_{j=1}^N \langle \bs{\Psi}_l,\bs{\Omega}_j^-\rangle\langle \bs{\Omega}_j^-, U^n\bs{\Psi}_m \rangle
\label{5eqn: in proof 5.2}
\end{align}
The first term becomes 
\begin{equation}\label{5eqn:in proof 5.2 101}
\langle \bs{\Psi}_l,\bs{\Omega}_0\rangle\langle \bs{\Omega}_0, U^n\bs{\Psi}_m \rangle
=\langle \bs{\Psi}_l,\bs{\Omega}_0\rangle\langle U^{-n}\bs{\Omega}_0, \bs{\Psi}_m \rangle
=\langle \bs{\Psi}_l,\bs{\Omega}_0\rangle\langle \bs{\Omega}_0, \bs{\Psi}_m \rangle.
\end{equation}
For the second term of \eqref{5eqn: in proof 5.2} we see that
\begin{align*}
\sum_{j=1}^N \langle \bs{\Psi}_l,\bs{\Omega}_j^+\rangle\langle \bs{\Omega}_j^+, U^n\bs{\Psi}_m \rangle
&=\sum_{j=1}^N \langle \bs{\Psi}_l,\bs{\Omega}_j^+\rangle
 \langle U^{-n}\bs{\Omega}_j^+, \bs{\Psi}_m \rangle \\
&=\sum_{j=1}^N \langle \bs{\Psi}_l,\bs{\Omega}_j^+\rangle
 \langle e^{-in\theta_j}\bs{\Omega}_j^+, \bs{\Psi}_m \rangle \\
&=\sum_{j=1}^N e^{in\theta_j}\langle \bs{\Psi}_l,\bs{\Omega}_j^+\rangle
 \langle \bs{\Omega}_j^+, \bs{\Psi}_m \rangle.
\end{align*}
Then we apply Lemma \ref{4lem:4.33} to have
\begin{align}
\sum_{j=1}^N \langle \bs{\Psi}_l,\bs{\Omega}_j^+\rangle\langle \bs{\Omega}_j^+, U^n\bs{\Psi}_m \rangle
&=\sum_{j=1}^N e^{in\theta_j}
  \frac{- ie^{i\theta_j}}{\sqrt{2}}\langle \bs{\Psi}_l,\bs{\Omega}_j \rangle
  \frac{ie^{-i\theta_j}}{\sqrt{2}}\langle\bs{\Omega}_j, \bs{\Psi}_m \rangle 
\nonumber \\
&=\sum_{j=1}^N \frac{e^{in\theta_j}}{2}\,
  \langle \bs{\Psi}_l,\bs{\Omega}_j \rangle
  \langle\bs{\Omega}_j, \bs{\Psi}_m \rangle.
\label{5eqn:in proof 5.2 102}
\end{align}
Applying a similar argument to the third term of \eqref{5eqn: in proof 5.2}, we obtain 
\begin{equation}\label{5eqn:in proof 5.2 103}
\sum_{j=1}^N \langle \bs{\Psi}_l,\bs{\Omega}_j^-\rangle\langle \bs{\Omega}_j^-, U^n\bs{\Psi}_m \rangle
=\sum_{j=1}^N \frac{e^{-in\theta_j}}{2}\,
  \langle \bs{\Psi}_l,\bs{\Omega}_j \rangle
  \langle\bs{\Omega}_j, \bs{\Psi}_m \rangle.
\end{equation}
Summing up \eqref{5eqn:in proof 5.2 101}--\eqref{5eqn:in proof 5.2 103}, we see that
\eqref{5eqn: in proof 5.2} becomes
\begin{align*}
\langle \bs{\Psi}_l,U^n\bs{\Psi}_m \rangle
&=\langle \bs{\Psi}_l,\bs{\Omega}_0\rangle\langle \bs{\Omega}_0, \bs{\Psi}_m \rangle
+\sum_{j=1}^N \frac{e^{in\theta_j}+e^{-in\theta_j}}{2}\,
  \langle \bs{\Psi}_l,\bs{\Omega}_j \rangle
  \langle\bs{\Omega}_j, \bs{\Psi}_m \rangle \\
&=\sum_{j=0}^N (\cos n\theta_j)
  \langle \bs{\Psi}_l,\bs{\Omega}_j \rangle
  \langle\bs{\Omega}_j, \bs{\Psi}_m \rangle,
\end{align*}
where $\theta_0=0$ is taken into account.
Thus, \eqref{4eqn:5-2-01} is proved.

Noting that $U=SC$ and $C$ acts on $\Gamma(N)$ as the identity, we see that
\begin{align*}
\langle S\bs{\Psi}_l,U^n\bs{\Psi}_m \rangle
&=\langle S\bs{\Psi}_l, SCU^{n-1}\bs{\Psi}_m \rangle \\
&=\langle \bs{\Psi}_l, CU^{n-1}\bs{\Psi}_m \rangle \\
&=\langle C\bs{\Psi}_l, U^{n-1}\bs{\Psi}_m \rangle \\
&=\langle \bs{\Psi}_l, U^{n-1}\bs{\Psi}_m \rangle,
\end{align*}
which proves \eqref{4eqn:5-2-02}.
We next observe that
\[
\langle \bs{\Psi}_l,U^n S\bs{\Psi}_m \rangle
=\langle \bs{\Psi}_l,U^n SC\bs{\Psi}_m \rangle 
=\langle \bs{\Psi}_l,U^{n+1}\bs{\Psi}_m \rangle,
\]
which proves \eqref{4eqn:5-2-03}.
Finally, we see that
\[
\langle S\bs{\Psi}_l, U^n S\bs{\Psi}_m \rangle
=\langle SC\bs{\Psi}_l, U^n SC\bs{\Psi}_m \rangle
=\langle U\bs{\Psi}_l, U^{n+1}\bs{\Psi}_m \rangle
=\langle \bs{\Psi}_l,U^{n}\bs{\Psi}_m \rangle,
\]
which shows \eqref{4eqn:5-2-04}.
\begin{flushright}$\square$\end{flushright}

Let $\{P_n\,;\,n=0,1,2,\dots\}$ be the orthogonal polynomials with respect to the free Meixner law with
parameters $q,pq,r$, i.e.,
the polynomials defined by the Jacobi parameters
\[
\omega_1=q,
\,\,\,
\omega_2=\omega_3=\dots=pq;
\quad
\alpha_1=0,
\,\,\,
\alpha_2=\alpha_3=\dots=r,
\]
see also Appendix.
We set
\begin{align}
p_0(x)
&=P_0(x)=1, 
\nonumber \\
p_j(x)
&=\frac{P_j(x)}{\sqrt{\mathstrut\omega_1\dots \omega_j}}
 =\frac{P_j(x)}{\sqrt{q(pq)^{j-1}}}\,,
\quad j=1,2,\dots,N.
\label{5eqn:def of p_j}
\end{align}
It is shown that $\{p_j\,;\, 0\le j\le N\}$ satisfies the recurrence relations
determined by $T_N$.
We define 
\[
\mu_N=\sum_{j=0}^N \rho(j)\delta_{\lambda_j}\,,
\qquad
\rho(j)=\rho_N(j)=\bigg(\sum_{n=0}^N p_n(\lambda_j)^2\bigg)^{-1}.
\]
The following results are known by general theory of Jacobi matrices and orthogonal polynomials. \cite{Deift,HO}

\begin{lemma}
$\mu_N$ is a probability distribution uniquely determined by the Jacobi matrix $T_N$.
Moreover, $\{p_j\,;\,j=0,1,2,\dots,N\}$ is the orthogonal polynomials with respect to $\mu_N$, normalized 
so as to have norm one, i.e.,
\[
\int_{-1}^1 p_j(x)p_k(x)\mu_N(dx)=\delta_{jk}\,.
\]
\end{lemma}

\begin{lemma}\label{5lem:explicit form of Omega_j}
For $j=0,1,\dots,N$ 
let $\bs{\Omega}_j$ be the normalized eigenvector of $T_N$ with eigenvalue $\lambda_j$ such that
$\langle \bs{\Omega}_j, \bs{\Psi}_0\rangle>0$.
Then, 
\[
\bs{\Omega}_j=\sqrt{\rho_N(j)}\sum_{n=0}^N p_n(\lambda_j)\bs{\Psi}_n\,,
\]
or equivalently,
\[
\langle \bs{\Omega}_j, \bs{\Psi}_n\rangle
=\sqrt{\rho_N(j)}\, p_n(\lambda_j).
\]
\end{lemma}

The next result is a key for removing the cutoff.

\begin{lemma}\label{4lem:limit free Meixner law}
The sequence of probability distributions $\mu_N$ converges weakly to the
free Meixner law with parameters $q,pq,r$.
In particular, for any continuous function $f$ on $[-1,1]$ we have
\[
\lim_{N\rightarrow\infty} \int_{-1}^1 f(x)\mu_N(dx)= \int_{-1}^1 f(x)\mu(dx).
\]
\end{lemma}

\noindent{\it Proof.}
We first note that
\begin{equation}\label{eqn:convergence in moments}
\lim_{N\rightarrow\infty} \int_{-\infty}^{+\infty} x^m\mu_N(dx)= \int_{-1}^1 x^m\mu(dx),
\qquad m=0,1,2,\dots.
\end{equation}
In fact, the $m$-th moment of $\mu_N$ is a polynomial in the
first $m$ terms of the Jacobi coefficients of $\mu_N$,
which are identical with the first $m$ terms of the Jacobi coefficients of the free
Meixner law $\mu$ if $m<N$.
Since the free Meixner law has a compact support,
it is uniquely determined by the moment sequence.
Therefore, it follows by general theory  
that \eqref{eqn:convergence in moments} implies the weak convergence
of $\mu_N$ to $\mu$.
\begin{flushright}$\square$\end{flushright}

\begin{theorem}[Integral representation of transition amplitude]
\label{5thm:integral representation of four amplitudes}
Let $U$ be the $(p,q)$-quantum walk on $\mathbb{Z}_+$
and $\mu$ the free Meixner law with parameters $q,pq,r$.
For any $l,m\in\mathbb{Z}_+$ and $n=0,\pm1,\pm2,\dots$ it holds that
\begin{equation}\label{4eqn:5-2-11}
\langle \bs{\Psi}_l,U^n\bs{\Psi}_m \rangle
=\int_{-1}^1 (\cos n\theta)\,p_l(\lambda)p_m(\lambda)\,\mu(d\lambda),
\quad \cos\theta=\lambda.
\end{equation}
Moreover,
\begin{align*}
\langle S\bs{\Psi}_l,U^n\bs{\Psi}_m \rangle
&=\int_{-1}^1 (\cos (n-1)\theta)\,p_l(\lambda)p_m(\lambda)\,\mu(d\lambda),
\\
\langle \bs{\Psi}_l,U^n S\bs{\Psi}_m \rangle
&=\int_{-1}^1 (\cos (n+1)\theta)\,p_l(\lambda)p_m(\lambda)\,\mu(d\lambda),
\\
\langle S\bs{\Psi}_l, U^n S\bs{\Psi}_m \rangle
&=\int_{-1}^1 (\cos n\theta)\,p_l(\lambda)p_m(\lambda)\,\mu(d\lambda).
\end{align*}
\end{theorem}

\noindent{\it Proof.}
Since $\langle \bs{\Psi}_l,U^n\bs{\Psi}_m \rangle$ coincides with the
similar expression for $(p,q)$-quantum walk on the path of length $N> \min\{l+n, m+n\}$.
We take such a sufficiently large $N$.
By Lemmas \ref{5lem:four inner products} and \ref{5lem:explicit form of Omega_j} we have
\begin{align}
\langle \bs{\Psi}_l,U^n\bs{\Psi}_m \rangle
&=\sum_{j=0}^N (\cos n\theta_j)
 \langle \bs{\Psi}_l, \bs{\Omega}_j\rangle
 \langle \bs{\Omega}_j, \bs{\Psi}_m\rangle 
\nonumber \\
&=\sum_{j=0}^N (\cos n\theta_j)\, p_l(\lambda_j) p_m(\lambda_j) \rho(j) 
\nonumber \\
&=\int_{-1}^1 (\cos n\theta)\, p_l(\lambda)p_m(\lambda) \mu_N(d\lambda),
\end{align}
which holds for all sufficiently large $N$.
Then, taking Lemma \ref{4lem:limit free Meixner law} into account, we come to
\begin{align*}
\langle \bs{\Psi}_l,U^n\bs{\Psi}_m \rangle
&=\lim_{N\rightarrow\infty}\int_{-1}^1  (\cos n\theta)\,p_l(\lambda)p_m(\lambda) \mu_N(d\lambda) \\
&=\int_{-1}^1 (\cos n\theta)\, p_l(\lambda)p_m(\lambda) \mu(d\lambda).
\end{align*}
This completes the proof of \eqref{4eqn:5-2-11}.
The rest is proved by combination of
Lemma \ref{5lem:four inner products} and \eqref{4eqn:5-2-11}.
\begin{flushright}$\square$\end{flushright}

\begin{theorem}\label{5thm:localization criteria}
Let $U$ be the $(p,q)$-quantum walk on $\mathbb{Z}_+$ with parameters satisfying
\begin{equation}\label{5eqn:condition for pqr}
p+q+r=1,
\qquad p\ge q>0,
\qquad r\ge0.
\end{equation}
Then it holds that
\[
\langle \bs{\Psi}_l, U^n \bs{\Psi}_0\rangle
\sim
w p_l(\xi)\cos n \tilde\theta,
\qquad
\text{as $n\rightarrow\infty$},
\]
where
\[
w=\max\left\{\frac{(1-p)^2-pq}{(1-p)(1-p+q)}\,,0\right\},
\quad
\xi=-\frac{q}{1-p}=\cos\tilde\theta,
\quad
0<\tilde\theta <\pi.
\]
Therefore,
\[
\lim_{N\rightarrow\infty}\frac1N\sum_{n=0}^{N-1}
|\langle \bs{\Psi}_0, U^n \bs{\Psi}_0\rangle|^2
=\frac{w^2}{2}\,.
\]
In particular, $U$ exhibits the initial point localization 
if and only if $w>0$, i.e., $(1-p)^2-pq>0$.
\end{theorem}

\noindent{\it Proof.}
By \eqref{4eqn:5-2-11} we have
\begin{equation}\label{5eqn:in proof Theorem 5.7} 
\langle \bs{\Psi}_l,U^n\bs{\Psi}_0 \rangle
=\int_{-1}^1 (\cos n\theta)\,p_l(\lambda)\,\mu(d\lambda),
\quad \cos\theta=\lambda.
\end{equation}
Under the assumption \eqref{5eqn:condition for pqr}
the free Meixner law with parameters $q,pq,r$ is of the form:
\[
\mu(dx)=\rho(x)dx+w\delta_\xi \,,
\]
where $\rho$ is a continuous function on $[r-2\sqrt{pq}\,, r+2\sqrt{pq}\,]\subset [-1,1]$,
an explicit form is deferred in Appendix, and
\[
\xi=-\frac{q}{1-p}\,,
\quad
w=\max\left\{\frac{(1-p)^2-pq}{(1-p)(1-p+q)}\,,0\right\}.
\]
Since $\rho$ is an integrable function, the Riemann--Lebesgue lemma implies that
\[
\lim_{n\rightarrow\infty} \int_{-1}^{1} (\cos n\theta)\,p_l(\lambda) \rho(\lambda) \, d\lambda=0.
\]
Hence in \eqref{5eqn:in proof Theorem 5.7} only contribution by the point mass
remains in the limit, i.e.,
\[
\langle \bs{\Psi}_l, U^n \bs{\Psi}_0\rangle
\sim
w p_l(\xi) \cos n \tilde\theta,
\qquad
\text{as $n\rightarrow\infty$},
\]
as desired.
The rest is straightforward.
\begin{flushright}$\square$\end{flushright}

\subsection{Proof of Theorem \ref{mainthm:probability amplitude}}

Let $U$ be the Grover walk on a spidernet $G=S(a,b,c)$ and consider
the initial state $\bs{\psi}^+_0$ defined by \eqref{3eqn:initial state}.
Let
\[
\Gamma(\mathbb{Z}_+)\subset\mathcal{H}(\mathbb{Z}_+)\subset \mathcal{H}(G)
\]
be the subspaces defined in Subsection \ref{subsec:4.1}.
Then ${\mathcal{H}(\mathbb{Z}_+)}$ is invariant under $U$ and
$U\!\!\restriction_{\mathcal{H}(\mathbb{Z}_+)}$ is the $(p,q)$-quantum walk on $\mathbb{Z}_+$,
where
\begin{equation}\label{5eqn:pqr by abc}
p=\frac{c}{b}\,,
\qquad
q=\frac{1}{b}\,,
\qquad
r=\frac{b-c-1}{b}\,.
\end{equation}
Since the initial state $\bs{\psi}^+_0=\bs{\Psi}_0$ belongs to $\mathcal{H}(\mathbb{Z}_+)$,
$\langle \bs{\psi}_0^+, U^n \bs{\psi}_0^+\rangle$ is obtained from the
$(p,q)$-quantum walk on $\mathbb{Z}_+$.
In fact, by Theorem \ref{5thm:integral representation of four amplitudes}
we have
\begin{equation}\label{5eqn:main representation}
\langle \bs{\psi}_0^+, U^n \bs{\psi}_0^+\rangle
=\int_{-1}^{1} (\cos n\theta) \,\mu(d\lambda),
\quad
\lambda=\cos\theta,
\end{equation}
where $\mu$ be the free Meixner law with parameters $q,pq,r$.
This completes the proof of Theorem \ref{mainthm:probability amplitude}.

\subsection{Proof of Theorem \ref{3thm:localization for spidernet}}

For a spidernet $G=S(a,b,c)$ the parameters $p,q,r$ defined by
\eqref{5eqn:pqr by abc} satisfies the condition in
Theorem \ref{5thm:localization criteria}.
So it holds that
\[
\langle \bs{\Psi}_0, U^n \bs{\Psi}_0\rangle
\sim
w \cos n \tilde\theta,
\qquad
\text{as $n\rightarrow\infty$},
\]
where
\begin{align}
w&=\max\left\{\frac{(1-p)^2-pq}{(1-p)(1-p+q)}\,,0\right\},
\label{5eqn: theta and w} \\
\xi&=-\frac{q}{1-p}=\cos\tilde\theta,
\quad 0<\tilde\theta <\pi.
\label{5eqn: c and tilde theta}
\end{align}
For the first half of Theorem \ref{3thm:localization for spidernet} 
it is sufficient to apply the following obvious relations:
\[
\frac{(1-p)^2-pq}{(1-p)(1-p+q)}=\frac{(b-c)^2-c}{(b-c)(b-c+1)}\,,
\qquad
-\frac{q}{1-p}=-\frac{1}{b-c}\,.
\]
For the second half we need only to note that
$(b-c)^2-c>0$ is equivalent to $b>c+\sqrt{c}$ under the assumption
\eqref{3eqn:constraint abc} posed at the beginning.

\subsection{Proofs of Corollaries \ref{3cor:k<10}--\ref{tree}}

These follow immediately from Theorem \ref{3thm:localization for spidernet}.
We need only to check the parameters.
For a spidernet $S(\kappa,\kappa+2,\kappa-1)$ we have
\begin{align*}
\xi&=\cos\tilde\theta=-\frac{1}{b-c}=-\frac{1}{3}\,, \\
w&=\max\left\{\frac{(b-c)^2-c}{(b-c)(b-c+1)}\,,0\right\}
=\max\left\{\frac{10-\kappa}{12}\,,0\right\}.
\end{align*}
While, for a spidernet $S(a,b,b-1)$ we have
\[
(b-c)^2-c=-(b-2)\le 0,
\qquad \kappa\ge2,
\]
which implies $w=0$.

\subsection{Proof of Theorem \ref{local}}

In a similar manner as in the proof of Theorem \ref{3thm:localization for spidernet}
we see that
\begin{equation}\label{4eqn:5-2-110}
\langle \bs{\Psi}_l,U^n\bs{\Psi}_0 \rangle
\sim w p_l(\xi) \cos n\tilde\theta,
\quad\text{as $n\rightarrow\infty$},
\end{equation}
where $w$, $\xi$, $\tilde\theta$ are given by \eqref{5eqn: theta and w} and \eqref{5eqn: c and tilde theta}.
The value $p_l(\xi)$ is known explicitly from Lemma \eqref{5lem:special value} below:
\[
p_l(\xi)=\frac{1}{\sqrt{p}}
\bigg(-\frac{\sqrt{\mathstrut pq}}{1-p}\bigg)^l
=\sqrt{\frac{b}{c}}\left(-\frac{\sqrt{c}}{b-c}\right)^l,
\qquad l=1,2,\dots.
\]
Then the time averaged limit probability is given by
\begin{align}
&\lim_{N\rightarrow\infty}\frac{1}{N}\sum_{n=0}^{N-1}
 |\langle \bs{\Psi}_l, U^n\bs{\Psi}_0 \rangle |^2
=\frac{w^2}{2}p_l(\xi)^2
\nonumber \\
&\qquad
=\frac12\left\{\frac{(b-c)^2-c}{(b-c)(b-c+1)}\right\}^2
 \times \frac{b}{c} \left\{\frac{c}{(b-c)^2}\right\}^l.
\label{exponential(0)} 
\end{align}
We here use the following rather obvious result.

\begin{lemma}\label{5lem:5.7}
Let $U$ be the Grover walk on a spidernet $S(a,b,c)$ with an initial state $\bs{\psi}_0^+$.
Then we have
\begin{equation}\label{5eqn:in lem 5.7(0)}
P(X_n\in V_l)
\ge|\langle \bs{\Psi}_l, U^n\bs{\Psi}_0\rangle|^2,
\qquad l=1,2,\dots.
\end{equation}
\end{lemma}

\noindent{\it Proof.}
We first note the obvious inequality:
\begin{align}
P(X_n\in V_l)
&=\sum_{u\in V_l} \sum_{v\sim u} 
 |\langle \bs{\delta}_{(u,v)}, U^n\bs{\psi}_0^+\rangle|^2 
\nonumber \\
&\ge \left|\left\langle  \frac{1}{\sqrt{b |V_l|}}\sum_{u\in V_l} \sum_{v\sim u} 
  \bs{\delta}_{(u,v)}, U^n\bs{\Psi}_0\right\rangle\right|^2.
\label{5eqn:in proof 5.7}
\end{align}
On the other hand, from the definitions \eqref{3eqn:def of psi+}--\eqref{3eqn:def of psi-} we see that
\[
\sum_{u\in V_l} \sum_{v\sim u} \bs{\delta}_{(u,v)}
=\sqrt{\mathstrut ac^{l}}\,\bs{\psi}_l^+ 
 +\sqrt{\mathstrut a(b-c-1)c^{l-1}}\,\bs{\psi}_l^\circ
  +\sqrt{\mathstrut ac^{l-1}}\, \bs{\psi}_l^-.
\]
Then, noting that $|V_l|=ac^{l-1}$ for $l\ge1$, we obtain
\[
\frac{1}{\sqrt{b |V_l|}}\sum_{u\in V_l} \sum_{v\sim u} \bs{\delta}_{(u,v)}
=\sqrt{p}\,\bs{\psi}_l^+ +\sqrt{r}\,\bs{\psi}_l^\circ+\sqrt{q}\, \bs{\psi}_l^-
=\bs{\Psi}_l\,.
\]
Inserting the above relation into \eqref{5eqn:in proof 5.7},
we obtain \eqref{5eqn:in lem 5.7(0)}.
\begin{flushright}$\square$\end{flushright}

Applying Lemma \ref{5lem:5.7} to \eqref{exponential(0)}, we obtain
\begin{equation}\label{5eqn:1 st of theorem}
\liminf_{N\rightarrow\infty}\frac{1}{N}\sum_{n=0}^{N-1}P(X_n\in V_l)
\ge \frac{b}{2c}\left\{\frac{(b-c)^2-c}{(b-c)(b-c+1)}\right\}^2
 \left\{\frac{c}{(b-c)^2}\right\}^l,
\end{equation}
which proves the first half of Theorem \ref{exponential localization}.
If the spidernet $S(a,b,c)$ is rotationally symmetric around $o$, we have
\[
P(X_n=u)
=\frac{1}{|V_l|}\, P(X_n\in V_l),
\qquad \partial(o,u)=l.
\]
Then the second half of Theorem \ref{exponential localization} follows by
dividing \eqref{5eqn:1 st of theorem} by $|V_l|=ab^{l-1}$.

Finally, we calculate the value of $p_l(x)$ at $x=\xi$.
The result is somehow amazing and plays a key role in showing the exponential localization.

\begin{lemma}\label{5lem:special value}
Let $p,q,r$ be constant numbers satisfying
\[
p>0,\quad
q>0, \quad
r=1-p-q\ge0, \quad
(1-p)^2-pq>0.
\]
Let $\{p_n\}$ be the orthogonal polynomials 
associated with the free Meixner law with parameters $q,pq,r$,
normalized to have norm one as before, see \eqref{5eqn:def of p_j}.
Then we have
\[
p_n\left(-\frac{q}{1-p}\right)=\frac{1}{\sqrt{p}}
\left(-\frac{\sqrt{\mathstrut pq}}{1-p}\right)^n,
\quad n=1,2,\dots.
\]
\end{lemma}

\noindent{\it Proof.}
We see from Theorem \ref{Athm: OP for free Meixner} that
the orthogonal polynomials $\{P_n\}$ associated with the free Meixner law with parameters $q,pq,r$
verify
\begin{equation}\label{5eqn:in 5.9 (1)}
P_n(x)=\frac{(xR_+(x)-2q)R_+(x)^{n-1}-(xR_-(x)-2q)R_-(x)^{n-1}}{2^{n-1}(R_+(x)-R_-(x))}\,,
\quad n\ge1,
\end{equation}
where
\[
R_{\pm}(x)=x-r \pm\sqrt{\mathstrut (x-r)^2-4pq}\,,
\quad
(x-r)^2-4pq>0.
\]
We need to compute the value of $P_n(x)$ at $\xi=-q/(1-p)$. 
Noting first that
\[
(\xi-r)^2-4pq
=\left(-\frac{q}{1-p}-r\right)^2-4pq
=\left\{\frac{(1-p)^2-pq}{1-p}\right\}^2,
\]
we obtain
\[
R_+(\xi)=-\frac{2pq}{1-p}\,,
\qquad
R_-(\xi)=-2(1-p),
\]
and hence 
\begin{gather*}
\xi R_+(\xi)-2q=\frac{2q(pq-(1-p)^2)}{(1-p)^2}\,,
\qquad
\xi R_-(\xi)-2q=0, \\
\xi(R_+(\xi)-R_-(\xi))=\xi R_+(\xi)-2q.
\end{gather*}
Then putting $x=\xi$ in \eqref{5eqn:in 5.9 (1)} we have
\begin{align*}
P_n(\xi)
&=\frac{(\xi R_+(\xi)-2q)R_+(\xi)^{n-1}}{2^{n-1}(R_+(\xi)-R_-(\xi))} \\
&=\frac{\xi}{2^{n-1}}\,R_+(\xi)^{n-1} \\
&=\frac{1}{p}\left(-\frac{pq}{1-p}\right)^n,
\qquad n=1,2,\dots.
\end{align*}
Finally, in view of \eqref{5eqn:def of p_j} we have
\[
p_n(x)
=\frac{P_n(x)}{\sqrt{\mathstrut q(pq)^{n-1}}}
=\frac{1}{\sqrt{\mathstrut p}}\left(-\frac{\sqrt{\mathstrut pq}}{1-p}\right)^n,
\quad n=1,2,\dots.
\]
This completes the proof.
\begin{flushright}$\square$\end{flushright}
\par
\par\noindent
\noindent
{\bf Acknowledgments.}
NK was partially supported by the Grant-in-Aid for Scientific Research (C) of Japan Society for the 
Promotion of Science (Grant No. 21540118). 
NO was partially supported by the CREST project ``A Mathematical Challenge to a New Phase of Material Sciences" (2008--2014)
of Japan Science and Technology Agency.
\par
\
\par

\begin{small}
\bibliographystyle{jplain}

\begin{thebibliography}{99}
\bibitem{Ahlbrecht}
A. Ahlbrecht, V. B. Scholz and A. H. Werner:
Disordered quantum walks in one lattice dimension, 
J. Math. Phys. {\bf 52} (2011), 102201. 

\bibitem{ABNVW}
A. Ambainis, E. Bach, A. Nayak, A. Vishwanath, and J. Watrous: 
One-dimensional quantum walks,
Proc. 33rd Annual ACM Symp. Theory of Computing, (2001) 37--49. 

\bibitem{AKR}
A. Ambainis, J. Kempe, A. Rivosh: 
Coins make quantum walks faster, 
Proc. 16th ACM-SIAM SODA (2005), 1099--1108.

\bibitem{Ambainis}
A. Ambainis: 
Quantum walks and their algorithmic applications, 
Int. J. Quantum Inf. {\bf 1} (2003), 507--518.  

\bibitem{Anshelevich}
M. Anshelevich:
Free Meixner states. Commun. Math. Phys. \textbf{276} (2007), 863--899. 

\bibitem{BB}
M. Bo\.zejko and W. Bryc:
On a class of free L{\'e}vy laws related to regression problem, 
J. Funct. Anal. {\bf 236} (2006), 59--77.


\bibitem{CGMV}
M. J. Cantero, F. A. Gr\"unbaum, L. Moral and L. Vel\'azquez: 
Matrix-valued Szeg\H{o} polynomials and quantum random walks, 
Commun. Pure Appl. Math. {\bf 63} (2010), 464--507.

\bibitem{CGMV2}
M. J. Cantero, F. A. Gr\"unbaum, L. Moral and L. Velazquez: 
One-dimensional quantum walks with one defect, 
Rev. Math. Phys. {\bf 24} (2012) 1250002. 

\bibitem{Chihara}
T. S. Chihara:
An Introduction to Orthogonal Polynomials,
Dover, New York, 2011.

\bibitem{CHKS}
K. Chisaki, M. Hamada, N. Konno and E. Segawa:
Limit theorems for discrete-time quantum walks on trees,
Interdiscip. Inform. Sci. {\bf 15} (2009), 423--429.

\bibitem{Deift}
P. A. Deift:
Orthogonal Polynomials and Random Matrices: 
A Riemann-Hilbert Approach,
Courant Lecture Notes in Mathematics Vol. 3,
American Mathematical Society, Providence, RI, 1999.


\bibitem{GS2006}
S. Gnutzmann and U. Smilansky:
Quantum graphs: Applications to quantum chaos and universal spectral statistics,
Adv. Phys. {\bfseries 55} (2006), 527--625.

\bibitem{Gudder}
S. P. Gudder: Quantum Probability, 
Academic Press, 1988.

\bibitem{GVWW}
F. A. Gr\"unbaum, L. Vel\'azquez, A. H. Werner, R. F. Werner: 
Recurrence for discrete time unitary evolutions, 
arXiv:1202.3903 (2012).

\bibitem{HO}
A. Hora and N. Obata:
Quantum Probability and Spectral Analysis of Graphs, 
Springer, 2007.

\bibitem{IO}
D. Igarashi and N. Obata:
Asymptotic spectral analysis of growing graphs: Odd graphs and spidernets,
Banach Center Publications {\bfseries 73} (2006), 245--265.

\bibitem{IKS}
N. Inui, N. Konno and E. Segawa:
One-dimensional three-state quantum walk, 
Physical Review E \textbf{72} (2005) 056112.

\bibitem{Joye}
A. Joye and M. Merkli:
Dynamical localization of quantum walks in random environments, 
J. Stat. Phys. \textbf{140} (2010), 1023--1053. 

\bibitem{KFCSWAMW}
M. Karski, L. F\"oster, J.-M. Choi, A. Steffen, W. Alt, D. Meschede and A. Widera: 
Quantum walk in position space with single optically trapped atoms, 
Science {\bf 325} (2009), 174--177.

\bibitem{Kesten}
H. Kesten:
Symmetric random walks on groups,
Trans. Amer. Math. Soc. \textbf{92} (1959), 336--354.

\bibitem{Konno2002}
N. Konno: Quantum random walks in one dimension,
Quantum Inf. Proc. {\bf 1} (2002), 345--354.

\bibitem{Konno2005}
N. Konno: A new type of limit theorems for the one-dimensional quantum random walk,
J. Math. Soc. Japan {\bfseries 57} (2005), 1179--1195.

\bibitem{KonnoBook}
N. Konno: 
Quantum walks, 
in ``Quantum Potential Theory
(U. Franz and M. Sch\"urmann, Eds.),"
pp. 309--452,
Lecture Notes in Math. {\bf 1954}, Springer, 2008. 

\bibitem{KLS}
N. Konno, T. {\L}uczak and E. Segawa:
Limit measures of inhomogeneous discrete-time quantum walks in one dimension, 
Quantum Inf. Proc., in press, arXiv:1107.4462 (2011). 

\bibitem{KS}
N. Konno and E. Segawa:
Localization of discrete-time quantum walks on a half line via the CGMV method,
Quantum Information and Computation {\bf 11} (2011), 0485--0495.

\bibitem{Meyer}
D. A. Meyer: 
From quantum cellular automata to quantum lattice gases, 
J. Stat. Phys. {\bf 85} (1996), 551--574.

\bibitem{Obata2012}
N. Obata:
One-mode interacting Fock spaces and random walks on graphs,
Stochastics {\bfseries 84} (2012), 383--392.

\bibitem{SY2001}
N. Saitoh and H. Yoshida:
The infinite divisibility and orthogonal polynomials with a constant
recursion formula in free probability theory,
Probab. Math. Statist. \textbf{21} (2001), 159--170.

\bibitem{S}
E. Segawa: 
Localization of quantum walks induced by recurrence properties of random walks, 
arXiv:1112.4982 (2011). 

\bibitem{SK}
Y. Shikano and H. Katsura:
Localization and factuality in inhomogeneous quantum walks with self-duality,
Phys. Rev. E \textbf{82} 031122 (2010).

\bibitem{TS}
T. Sunada and T. Tate: 
Asymptotic behavior of quantum walks on the line, 
J. Funct. Anal. {\bf 262} (2012), 2608--2645.

\bibitem{Sze}
M. Szegedy:
Quantum speed-up of Markov chain based algorithms, 
Proc. 45th Annual IEEE Symposium on Foundations of Computer Science (FOCS'04) (2004), 32--41.

\bibitem{Urakawa2003}
H. Urakawa: 
The Cheeger constant, the heat kernel, and the Green kernel of an infinite graph,
Monatsh. Math. \textbf{138} (2003), 225--237.

\bibitem{Venegas}
S. E. Venegas-Andraca: 
Quantum Walks for Computer Scientists, 
Morgan and Claypool, San Rafael, 2008.

\bibitem{Katori}
K. Watabe, N. Kobayashi, M. Katori and N. Konno: 
Limit distributions of two-dimensional quantum walks, 
Phys. Rev. A {\bf 77} (2008), 062331. 

\bibitem{Watrous}
J. Watrous: 
Quantum simulations of classical random walks and undirected graph connectivity, 
J. Comput. System Sci. {\bf 62} (2001), 376--391. 

\end{thebibliography}

\end{small}

\noindent\\
\noindent\\
\noindent\\

\noindent {\large{\bf Appendix A}: Free Meixner laws}  \\

\renewcommand{\thesection}{\Alph{section}}
\setcounter{section}{1}
\setcounter{theorem}{0}

\noindent The free Meixner law with parameters $p>0$, $q\ge0$, $a\in\mathbb{R}$ is a probability distribution $\mu$ 
on $\mathbb{R}$ uniquely determined by
\[
\int_{-\infty}^{+\infty} \frac{\mu(dx)}{z-x}
=\frac{1}{z}{\genfrac{}{}{0pt}{}{}{-}} 
\frac{p}{z-a}{\genfrac{}{}{0pt}{}{}{-}} 
\frac{q}{z-a}{\genfrac{}{}{0pt}{}{}{-}}
\frac{q}{z-a}{\genfrac{}{}{0pt}{}{}{-\dotsb}}\,,
\]
where the continued fraction in the right-hand side converges in $\mathbb{C}\backslash\mathbb{R}$.
In other words, $\mu$ is uniquely determined by the so-called Jacobi coefficients:
\[
\omega_1=p,\,\, \omega_2=\omega_3=\dots=q;
\qquad
\alpha_1=0, \,\, \alpha_2=\alpha_3=\dots=a.
\]
The free Meixner law with parameters $p=q=1$, $a=0$ is nothing else
the (normalized) \textit{Wigner semicircle law} and
the one with parameters $p>0$, $q\ge0$, $a=0$ the \textit{Kesten distribution}
\cite{Kesten} with parameters $p,q$.
The free Meixner laws have been studied mostly in 
the context of free probability and quantum probability \cite{Anshelevich,BB,HO,IO,SY2001}.

In general, with Jacobi parameters $\{\omega_n\}$, $\{\alpha_n\}$ we associate
a sequence of polynomials $\{P_n\}$ by
\begin{align*}
&P_0(x)=1, \\
&P_1(x)=x-\alpha_1, \\
&P_n(x)=P_{n+1}(x)+\alpha_{n+1}P_n(x)+\omega_n P_{n-1}(x),
\quad n\ge1.
\end{align*}
It is known that $\{P_n(x)\}$ is the orthogonal polynomials with respect to $\mu$.
We will derive an explicit expression of the orthogonal polynomials 
with respect to the free Meixner law.

Let $\{U_n\}$ be 
the Chebyshev polynomials of the second kind, i.e., defined by
\[
U_n(\cos\theta)=\frac{\sin(n+1)\theta}{\sin\theta}\,,
\qquad n=0,1,2,\dots.
\]
Set 
\[
\Tilde{U}_n(x)=U_n\Big(\frac{x}{2}\Big).
\]
It is well known that $\{\Tilde{U}_n\}$ form the orthogonal polynomials with
respect to the normalized Wigner semicircle law 
(the free Meixner law with parameters $p=q=1$, $a=0$) and 
are specified uniquely by the recurrence relations:
\begin{align*}
\Tilde{U}_0(x)&=1, \\
\Tilde{U}_1(x)&=x, \\
x\Tilde{U}_n(x)&=\Tilde{U}_{n+1}(x)+\Tilde{U}_{n-1}(x),
\qquad n\ge1.
\end{align*}

\begin{theorem}\label{athm:chebychev}
Let $p>0$, $q>0$ and $a\in\mathbb{R}$.
The orthogonal polynomial with respect to the
free Meixner law with parameters $p,q,a$ is given by
\begin{align*}
P_0(x)&=1, \\
P_1(x)&=x, \\
P_n(x)&=q^{n/2} \Tilde{U}_n\Big(\frac{x-a}{\sqrt{q}}\Big)
 +aq^{(n-1)/2} \Tilde{U}_{n-1}\Big(\frac{x-a}{\sqrt{q}}\Big) \\
&\qquad\qquad\qquad\qquad +(q-p)q^{(n-2)/2} \Tilde{U}_{n-2}\Big(\frac{x-a}{\sqrt{q}}\Big),
 \qquad n\ge2,
\end{align*}
\end{theorem}

\noindent{\it Proof.}
Set
\begin{align*}
V_0(x)&=1, \\
V_1(x)&=x, \\
V_n(x)&=q^{n/2}\left\{\Tilde{U}_n\Big(\frac{x}{\sqrt{q}}\Big)
 +\Big(1-\frac{p}{q}\Big)\Tilde{U}_{n-2}\Big(\frac{x}{\sqrt{q}}\Big)\right\},
\quad n\ge2.
\end{align*}
Then $V_n(x)=x^n+\dots$ and it holds that 
\begin{align*}
xV_1(x)&=V_2(x)+pV_0(x), \\
xV_n(x)&=V_{n+1}(x)+qV_{n-1}(x), \quad n\ge2.
\end{align*}
In other words, $\{V_n\}$ is the orthogonal polynomials with respect to the
Kesten distribution with parameters $p,q$. 
Then, it is straightforward to verify that the polynomials $\{P_n\}$ defined by
\begin{align*}
P_0(x)&=1, \\
P_n(x)&=V_n(x-a)+aq^{(n-1)/2}\Tilde{U}_{n-1}\Big(\frac{x-a}{\sqrt{q}}\Big),
\quad n\ge1,
\end{align*}
satisfy
\begin{align*}
P_0(x)&=1, \\
P_1(x)&=x, \\
xP_1(x)&=P_2(x)+aP_1(x)+pP_0(x), \\
xP_n(x)&=P_{n+1}(x)+aP_{n}(x)+qP_{n-1}(x), \quad n\ge2.
\end{align*}
This means that $\{P_n\}$ is the orthogonal polynomials with respect to 
the free Meixner law with parameters $p,q,a$.
\begin{flushright}$\square$\end{flushright}

By direct application of the famous expression of the Chebyshev polynomials of the second kind: 
\[
U_n(x)=\frac{\left(x+\sqrt{x^2-1}\,\right)^{n+1}-\left(x-\sqrt{x^2-1}\,\right)^{n+1}}
{2\sqrt{\mathstrut x^2-1}}\,,
\qquad n=0,1,2,\dots,
\]
which is valid for $|x|>1$,
we obtain a variant of Theorem \ref{athm:chebychev} as follows.

\begin{theorem}\label{Athm: OP for free Meixner}
Let $p>0$, $q>0$ and $a\in\mathbb{R}$.
The orthogonal polynomial with respect to the
free Meixner law with parameters $p,q,a$ is given by
\begin{align*}
P_0(x)&=1, \\
P_n(x)&=\frac{(xR_+(x)-2p)R_+(x)^{n-1}-(xR_-(x)-2p)R_-(x)^{n-1}}{2^{n-1}(R_+(x)-R_-(x))}\,,
\quad n\ge1,
\end{align*}
where
\[
R_{\pm}(x)=x-a \pm\sqrt{\mathstrut (x-a)^2-4q}\,,
\quad
(x-a)^2-4q>0.
\]
\end{theorem}

Finally, we mention briefly the explicit form of the free Meixner law.
For $p>0$, $q\ge0$, $a\in\mathbb{R}$ we set
\[
\rho(x)=\frac{p}{2\pi}\,\frac{\sqrt{4q-(x-a)^2}}{(q-p)x^2+pax+p^2}\,,
\qquad |x-a|\le 2\sqrt{q}\,.
\]
The free Meixner law is the sum of $\rho(x)dx$ and at most two atoms:
\[
\mu(dx)=\rho(x)dx+w_1\delta_{\xi_1}+w_2\delta_{\xi_2}\,,
\]
where $w_1\ge0$, $w_2\ge0$ and $\xi_1\neq \xi_2$.
For the explicit form, see e.g., \cite{HO,SY2001}.

\end{document}